\documentclass[useAMS,usenatbib,usegraphicx]{mn2e}
\usepackage{graphicx}                                            
\usepackage{times}
\usepackage{float}
\usepackage{rotating}
\usepackage{epstopdf}
\usepackage{multirow}
\usepackage{pdflscape}

\newcommand{\ergps}{erg~s$^{-1}$}

\def\ltsima{$\; \buildrel < \over \sim \;$}
\def\simlt{\lower.5ex\hbox{\ltsima}}
\def\gtsima{$\; \buildrel > \over \sim \;$}
\def\simgt{\lower.5ex\hbox{\gtsima}}
\def\gsimeq
{\hbox{\raise0.5ex\hbox{$>\lower1.06ex\hbox{$\kern-1.07em{\sim}$}$}}}
\def\lsimeq
{\hbox{\raise0.5ex\hbox{$<\lower1.06ex\hbox{$\kern-1.07em{\sim}$}$}}}

\def\xmm{{\it XMM-Newton }}

\def\asca{{\it ASCA}}

\def\xmm{{\it XMM-Newton}}
\def\chandra{{\it Chandra}}

\def\nustar{{\it NuSTAR}}
\def\swift{{\it Swift}}

\def\gmrt{{GMRT}}
\def\grond{{GROND}}

\def\apj{ApJ}
\def\mnras{MNRAS}
\def\aap{A\&A}
\def\apjl{ApJ}
\def\apjs{ApJS}
\def\araa{ARA\&A}
\def\pasj{PASJ}
\def\nat{Nature}

\def\ssr{SSRv}
\def\aapr{A\&ARv}
\def\aaps{A\&AS} 
\def\physrep{Phys. Rep.}
\def\pasp{PASP}

\def\axj{AX~J1745.6-2901}
\def\sgr{SGR~J1745-2900}
\def\sgras{Sgr~A$^{\star}$}
\def\swiftj{Swift~J174540.7-290015}

\def\xspec{{\it Xspec}}

\def\xis{XIS}
\def\xis1{XIS1}
\def\xis2{XIS2}
\def\xis3{XIS3}

\title[] 
 {{\nustar\ + \xmm\ monitoring of the neutron star transient \axj}}
 \author[G.\ Ponti et al. ]
 {G.~Ponti$^{1}$, S. Bianchi$^{2}$, T. Mu\~nos-Darias$^{3,4}$, K. Mori$^{5}$,  K. De$^{6}$, 
 A. Rau$^{1}$, B. De Marco$^{7}$, 
    \newauthor
C. Hailey$^{5}$,  J. Tomsick$^{8}$, K. K. Madsen$^9$, M. Clavel$^{8}$, F. Rahoui$^{10,11}$, D. V. Lal$^{12}$, 
    \newauthor
S. Roy$^{12}$ and  D.~Stern$^{13}$ 
 \\
   $^1$ Max-Planck-Institut f{\"u}r Extraterrestrische Physik, Giessenbachstrasse, D-85748, Garching, Germany\\
   $^2$ Dipartimento di Matematica e Fisica, Universit\`a Roma Tre, Via della Vasca Navale 84, I-00146, Roma, Italy\\
   $^{3}$ Instituto de Astrof\'isica de Canarias, 38205 La Laguna, Tenerife, Spain\\  
   $^{4}$ Departamento de astrof\'isica, Univ. de La Laguna, E-38206 La Laguna, Tenerife, Spain\\
   $^5$ Columbia Astrophysics Laboratory, Columbia University, New York, NY 10027, USA\\
   $^6$ California Institute of Technology, 1200, E. California Blvd., Pasadena, CA 91125, USA\\   
   $^7$ Nicolaus Copernicus Astronomical Center, Polish Academy of Sciences, Bartycka 18, PL-00-716 Warsaw, Poland \\
   $^8$ Space Sciences Laboratory, University of California, Berkeley, CA 94720, USA\\
   $^{9}$ Cahill Centre for Astronomy and Astrophysics, California Institute of Technology, Pasadena, CA 91125, USA\\
   $^{10}$ European Southern Observatory, K. Schwarzschild-Str. 2, 85748 Garching bei MŸnchen, Germany \\
   $^{11}$ Department of Astronomy, Harvard University, 60 Garden street, Cambridge, MA 02138, USA \\
   $^{12}$ National Centre for Radio Astrophysics (TIFR), Post Box 3, Ganeshkhind P.O., Pune 411 007, India\\
   $^{13}$ Jet Propulsion Laboratory, California Institute of Technology, Pasadena, CA 91109, USA\\ 
      }
\pagerange{\pageref{firstpage}--\pageref{lastpage}}
\usepackage{times}
\begin{document}
\label{firstpage}
 \maketitle
\begin{abstract}
\axj\ is a high-inclination (eclipsing) transient neutron star (NS) Low Mass X-ray
Binary (LMXB) showcasing intense ionised Fe~K absorption. 
We present here the analysis of 11 \xmm\ and 15 \nustar\ new data-sets 
(obtained between 2013-2016), therefore tripling the number of observations 
of \axj\ in outburst. 
Thanks to simultaneous \xmm\ and \nustar\ spectra, we greatly improve 
on the fitting of the X-ray continuum. 
During the soft state the emission can be described by a disk black body 
($kT\sim1.1-1.2$~keV and inner disc radius $r_{DBB}\sim14$~km), plus hot 
($kT\sim2.2-3.0$~keV) black body radiation with a small emitting 
radius ($r_{BB}\sim0.5-0.8$~km) likely associated with the boundary layer
or NS surface, plus a faint Comptonisation component. 
Imprinted on the spectra are clear absorption features created by both 
neutral and ionised matter. Additionally, positive residuals suggestive 
of an emission Fe~K$\alpha$ disc 
line and consistent with relativistic ionised reflection are present during 
the soft state, while such residuals are not significant during the hard state. 
The hard state spectra are characterised by a hard 
($\Gamma\sim1.9-2.1$) power law, showing no evidence for a high energy 
cut off ($kT_e>60-140$~keV) and implying a small optical depth ($\tau<1.6$). 
The new observations confirm the previously witnessed trend of exhibiting strong 
Fe~K absorption in the soft state, that significantly weakens during the hard state. 
Optical (\grond) and radio (\gmrt) observations suggest for \axj\ a standard 
broad band SED as typically observed in accreting neutron stars. 

\end{abstract}


\section{Introduction}

Accreting X-ray binaries are among the brightest objects of the X-ray sky 
(Voges et al. 1999). The source population is divided into several subclasses. 
Depending on the nature of the compact object, they are classified as either 
black holes (BH) or neutron stars (NS) X-ray binaries (Liu et al. 2000; 2001; 
Tetarenko et al. 2016; Corral-Santana et al. 2016). 
According to the type of the companion star, they are further separated into 
either high or low mass X-ray binaries (LMXB). The family of NS LMXB is 
additionally split into Atoll and Z-sources, specified by the shape of their 
colour-colour X-ray diagram (Hasinger \& van der Klis 1989). 
The former cover a lower and larger luminosity range 
($L\sim0.001-0.5$~L$_{Edd}$) and they typically show both a soft and 
a hard state (Mu\~{n}oz-Darias et al. 2014). In contrast, Z-sources 
radiate at luminosities close to Eddington and their spectra are 
usually soft showing a more complex phenomenology 
(Barret \& Olive 2002; Reig et al. 2004; van der Klis 2006). 

Despite the differences (such as a surface and its associated emission 
that is present in NS while it is absent in BH, consequently this induces 
a softer and lower temperature Comptonised component; Burke et al. 2017) 
all LMXB are linked through common features. 
It was recently shown that NS LMXB display spectral states and 
a hysteresis pattern similar to that observed in BH binaries, indicating 
a physical link between these different sources (Mu\~{n}oz-Darias et al. 2014). 
Similarly, both classes of sources display alike fast variability properties 
(Wijnands \& van der Klis 1999; Belloni, Psaltis \& van der Klis 2002; 
Belloni et al. 2011; Fender \& Mu\~{n}oz-Darias 2016; Motta et al. 2017). 
Regardless of these common behaviours, more complex 
spectral models have been typically applied to NS LMXB, compared 
to BH systems (see Barret et al. 2001 for a review). Indeed, the bulk of 
the emission from BH systems can generally be reproduced by a soft 
disc black body plus a hard Comptonisation component (Done et al. 2007; 
Dunn et al. 2010; 2011; Plant et al. 2014). While a larger array of models 
and different approaches have been employed to explain the emission 
from NS LMXB, partly disguising the similarities (Mitsuda et al. 1984; 1989; 
White et al. 1988; Church \& Balucinska-Church 1995; 2001; 2004; 
Barret et al. 2000; Gierlinski \& Done 2002; Iaria et al. 2005). 

Mainly based on the lessons learned from the studies of BH systems, 
Lin et al. (2007; 2009) proposed a new model for NS composed of 
three main emission components. This model is composed by the same 
two components required to fit the spectra of BH binaries (disc black body 
plus Comptonisation), with the addition of a black body component to 
reproduce the radiation from the NS boundary layer. 
It was successfully applied to observations of 
several Atoll and Z-sources (Lin et al. 2007; 2009; 
Armas Padilla et al. 2017). In particular, the model allows fitting both the 
soft and hard state observations, obtaining best fit parameters similar 
to those observed in BH systems (besides the boundary layer). 

During the soft state, the emission is usually dominated by the thermal 
components, with characteristic temperatures 
of $kT\sim0.5-2.0$~keV (Barret et al. 2000; Oosterbroek et al. 2001; 
Di Salvo et al. 2000; Iaria et al. 2005; Lin et al. 2007; 2009; Armas Padilla 
et al. 2017). The Comptonised component is weak with low temperature 
($kT_e\sim$few tens keV) and large optical 
depth ($\tau\sim5-15$, for a spherical geometry). 
In the hard state, the spectra are dominated by a hard Comptonised 
component with temperatures of a few tens of keV and optical depths 
($\tau\sim2-3$). The thermal components are observed 
at low temperatures ($kT<1$~keV) and with significantly lower luminosity. 

Thanks to the combination of increased effective area and energy resolution, 
a number of additional narrow and broad features were detected 
in the recent years. Indeed, broad Fe K$\alpha$ emission lines 
are often measured (Miller et al. 2006; 2007; Done \& Gierlinski 2006; 
Reis et al. 2008; 2010; Yamada et al. 2009; Tomsick et al. 2009; 
Done \& Diaz-Trigo 2010; Shidatsu et al. 2011; Petrucci et al. 2014; 
Kolehmainen et al. 2014). The shape of such lines carries information 
about the geometry and the extension of the accretion disc as well as 
of the extent and location of the primary source (Fabian et al. 1989; Reynolds \& 
Nowak 2003; Fabian \& Ross 2010). 

Additionally, the past decade has seen a burst of detections of ionised 
absorption lines and edges in high inclination LMXB (Brandt et al. 2000; 
Lee et al. 2002; Parmar et al. 2002; Ueda et al. 2004; Miller et al. 2006a,b; 
Diaz-Trigo et al. 2006). It is now recognised that winds/ionised absorption 
are an ubiquitous component of accreting LMXB during the soft state 
(Ponti et al. 2012; 2016). 

The unprecedented focussing capability of the mirrors aboard
the \nustar\ telescope provides an improvement of the sensitivity in the 
10-79 keV energy range by more than two orders of magnitudes 
compared to previous coded mask telescopes (Harrison et al. 2013). 
This opens a new window for the study of both faint X-ray sources, 
such as the less luminous Atoll sources, as well as sources in crowded regions, 
such as the Galactic center (GC; Ponti et al. 2015). 

Looking into the \nustar\ and \xmm\ archives, we noted that \axj\ is one 
of the transient LMXB with the best \nustar+\xmm\ publicly available 
monitoring campaign. 

\subsection{\axj}

\axj\ is a high inclination (dipping and eclipsing) accreting NS transient, 
classified as an Atoll source and showing clear evidence for highly 
ionised absorption as well as type I bursts (Maeda et al. 1996; 
Hyodo et al. 2009; Ponti et al. 2015a). 
\axj\ is located towards the GC region at a projected distance of less 
than 1.5$^{\prime}$ from \sgras\ (Degenaar et al. 2009; 
Ponti et al. 2015b). The high column density of neutral material 
($N_H\sim3\times10^{23}$~cm$^{-2}$) as well as the spatial distribution
of the dust scattering halo suggest that \axj\ is either at or behind the GC, 
therefore at a distance $\geq 8$~kpc (Ponti et al. 2017; 
Jin et al. 2017). The source has an orbital period of 
$8.35100811\pm0.00000002$~hr decreasing with time at a rate 
of $\dot{P}_{orb}=-4.03\pm0.32\times10^{-11}$~s~s$^{-1}$,
suggesting non-conservative mass transfer (Maeda et al. 1996; Ponti et 
al. 2016b). 

Being located at less than 1.5$^{\prime}$ from \sgras, \axj\ is 
one of the best monitored accreting compact objects, having deep 
and frequent \xmm\ and \nustar\ observations (see Ponti et al. 2015b,c). 
\axj\ was discovered by \asca\ during the 1993-1994 outburst 
(Maeda et al. 1996). The source recently showed strong activity with 
two short ($\sim4-7$ months) outbursts in 2006 and 2007, as well as 
two long and luminous outbursts in 2007-2008 and 2013-2016 
(Degenaar et al. 2013; Ponti et al. 2015). The latest ended in June 
2016 (Degenaar et al. 2016). The intense activity together with the
frequent X-ray monitoring of this field make \axj\ one of the best 
candidates to investigate the evolution of the X-ray emission 
and constrain the spectral energy distribution (SED) in this low luminosity 
GC Atoll source. This will be instrumental to investigate, in a following 
paper (Bianchi et al. 2017), whether the ionised absorption observed 
towards \axj\ dissolves in the hard state because it becomes 
unstable for photo-ionisation. 

\

We report here 11 new \xmm\ observations (4 and 7 of which are in the soft and 
hard state, respectively) as well as 15 new \nustar\ observations that 
caught \axj\ in outburst. In particular we report the results of a set of 
five strictly simultaneous \xmm+\nustar\ observations, excellent for 
characterising the X-ray SED above 10~keV within the crowded GC 
region (Ponti et al. 2015b). 

\section{Observations and data reduction}

All spectral fits were performed using the \xspec\ software package 
(version 12.7.0; Arnaud 1996). 
Uncertainties and upper limits are reported at the 90 per cent confidence level for 
one interesting parameter, unless otherwise stated. 
The reported X-ray fluxes are not corrected for Galactic absorption. 
All luminosities, black body and disc black body radii assume that \axj\ 
is located at the GC distance, therefore $d_{AXJ}=8.3$~kpc (Genzel et 
al. 2010; Bland-Hawthorn et al. 2016; Gillessen et al. 2016). To derive 
the disc black body inner radius $r_{DBB}$, we fit the spectrum with the 
disc black body model (Mitsuda et al. 1984; Makishima et al. 1986)
the normalisation of which provides the apparent inner disc radius ($R_{DBB}$). 
Following Kubota et al. (1998), we correct the apparent inner 
disc radius through the equation: $r_{DBB}=\xi \kappa^2 R_{DBB}$ 
(where $\kappa=2$ and $\xi=\sqrt{(3/7)}\times(6/7)^3$) in order to estimate 
the likely inner disc radius $r_{DBB}$. We also assume an inclination of the 
accretion disc of $80^{\circ}$ (\axj\ is an eclipsing source). 
We adopt a nominal Eddington limit 
for \axj\ of  L$_{Edd}=2\times10^{38}$~\ergps (appropriate for a primary mass 
of $M_{NS}\sim1.4$~M$_{\odot}$ and cosmic composition; Lewin et al. 1993). 
To allow the use of $\chi^2$ statistics we group each spectrum to have 
a minimum of 30 counts in each bin. We fit the interstellar absorption 
with either the {\sc tbabs} or the {\sc tbvarabs} models in \xspec\ 
assuming Wilms et al. (2000) abundances and Verner et al. (1996) cross sections. 

\subsection{\xmm}
\label{xmmDR}

We followed the same data reduction steps (e.g., filtering out bursts 
and dipping periods) employed by Ponti et al. (2015a). 
We reanalysed all the data-sets considered in that work, with the latest version 
(15.0.0) of the \xmm\ (Jansen et al. 2001; Str\"{u}der et al. 2001; Turner et 
al. 2001) Science Analysis System {\sc sas}, applying the most 
recent (as of 2016 September 2) calibrations (see Tab. 1 of Ponti et al. 2015). 
We also report the analysis of 11 new \xmm\ observations of \axj\ in 
outburst (the study of other sources within the field of view of the same 
observations are included in Ponti et al. 2015b,c; 2016a,b,c). 
Tab. \ref{data} shows the details of all the new \xmm\ observations
considered in this work (see Tab. 1 of Ponti et al. 2015 for the old 
observations). All the new observations have been accumulated in Full Frame 
mode with the medium filter applied. Pile-up distorts the source spectra 
in the soft state. We therefore adopt an annular extraction region, while a 
circular one is considered in the hard state (see Ponti et al. 2015). 
We corrected the spectra of \axj\ for the effects of the dust scattering 
halo (see Jin et al. 2017a,b). 
During obsid 0790180401 the spectrum of \axj\ is affected by the 
emission from \swiftj, characterised by a very soft spectrum 
(see Ponti et al. 2016). To avoid possible biases, we decided to 
disregard this data set from further analysis. 

\subsection{\nustar}
\label{NuDR}

We analysed all \nustar\ (Harrison et al. 2013) public (as of 2016 September 2) 
observations that caught \axj\ in outburst, resulting in 15  separate data-sets 
(see Tab. \ref{TabNu}). 
We reduced and cleaned the data with the latest version of the \nustar\ CALDB 
(20161021). We used the NuSTARDAS sub-package v1.6.0 that is part of 
the HEASOFT v.6.19 (see Wik et al. 2014; Madsen et al. 2015 for more 
information about the calibration of the \nustar\ observatory). 
The \nustar\ data were reduced with the standard {\it nupipeline} 
scripts v. 0.4.5 (released on 2016-03-30) and the high level products produced 
with the {\it nuproducts} tool. 
The source photons were extracted from a circular region of $70^{\prime\prime}$ 
radius, centred on the source (Fig. \ref{NuIma}). 
Response matrices appropriate for each data-set were generated using 
the standard software. We did not combine modules. 

Bursts have been removed by generating a light curve with 3~s time 
binning in the 5-10 keV energy band and cutting all intervals with a count rate 
higher than the threshold reported in Tab. \ref{TabNu}. 
We also filtered out eclipses and dips by producing a light curve in the 
same energy band (with 60~s and 180~s time bins for the soft and 
hard state observations, respectively) and dismissing all periods with count 
rate lower than the threshold reported in Tab. \ref{TabNu}. 

In appendix \S~\ref{NDR}, we report more details on the treatment of the bright 
point sources contaminating the spectrum of \axj\ (\S \ref{Trans}), on the 
treatment of the background emission (\S \ref{Back}), as well as the 
potential mismatch at low energy between the spectrum measured 
by \nustar\ and by \xmm\ (\S \ref{Xcal}). 
As detailed in \S~A3, we limit the \nustar\ spectral analysis to 
the $5.5-40$ and $3-70$~keV energy range during the soft and hard state, 
respectively. 

\subsection{\swift}

Following the same procedure described in Ponti et al. (2015), we have 
extracted a \swift-XRT light curve of \axj\ from all 'photon counting' mode 
observations available as of 2016 September 2. 

\subsection{\grond}
\label{grond}

On 2015 April 29$^{\rm th}$ \axj\ was observed during the soft X-ray 
spectral state with the Gamma-Ray burst Optical Near-infrared 
Detector GROND (Greiner et al. 2008) at the MPG 2.2m telescope 
in La Silla, Chile. GROND provides simultaneous imaging in four optical 
(g$^{\prime}$, r$^{\prime}$, i$^{\prime}$, z$^{\prime}$) and three 
near-infrared (J, H, K$_s$) channels. Owing to the extreme Galactic 
foreground reddening towards the target (E$_{\rm B-V}\sim86$\,mag; 
Schlafly \& Finkbeiner 2011) only  the analysis of the near-IR bands 
is discussed here.

The data were reduced with the standard tools and methods described 
in Kr\"uhler et al. (2008). To search for a near-IR counterpart to \axj\ 
we combined the best seeing images taken during the night, providing 
total integration times of 94, 72, and 96\,min in J, H, and K$_s$, respectively. 
The resulting median full width at half maximum (FWHM) of the point 
spread function was $1.1^{\prime\prime}$, $1.6^{\prime\prime}$, and 
$1.2^{\prime\prime}$ in the three bands, respectively. 
The astrometric solution, with an accuracy of $0.01^{\prime\prime}$ 
in both coordinates, was obtained using selected 2MASS field 
stars (Skrutskie et al. 2006). 
The photometry was  measured from apertures with sizes corresponding 
to the image FWHM and calibrated against 2MASS field stars.  
This resulted in 1$\sigma$ systematic uncertainties of 
0.10\,mag (J), 0.09\,mag (H), and 0.17\,mag (K$_s$).  

\begin{figure}
\fbox{\includegraphics[height=0.22\textwidth,angle=0]{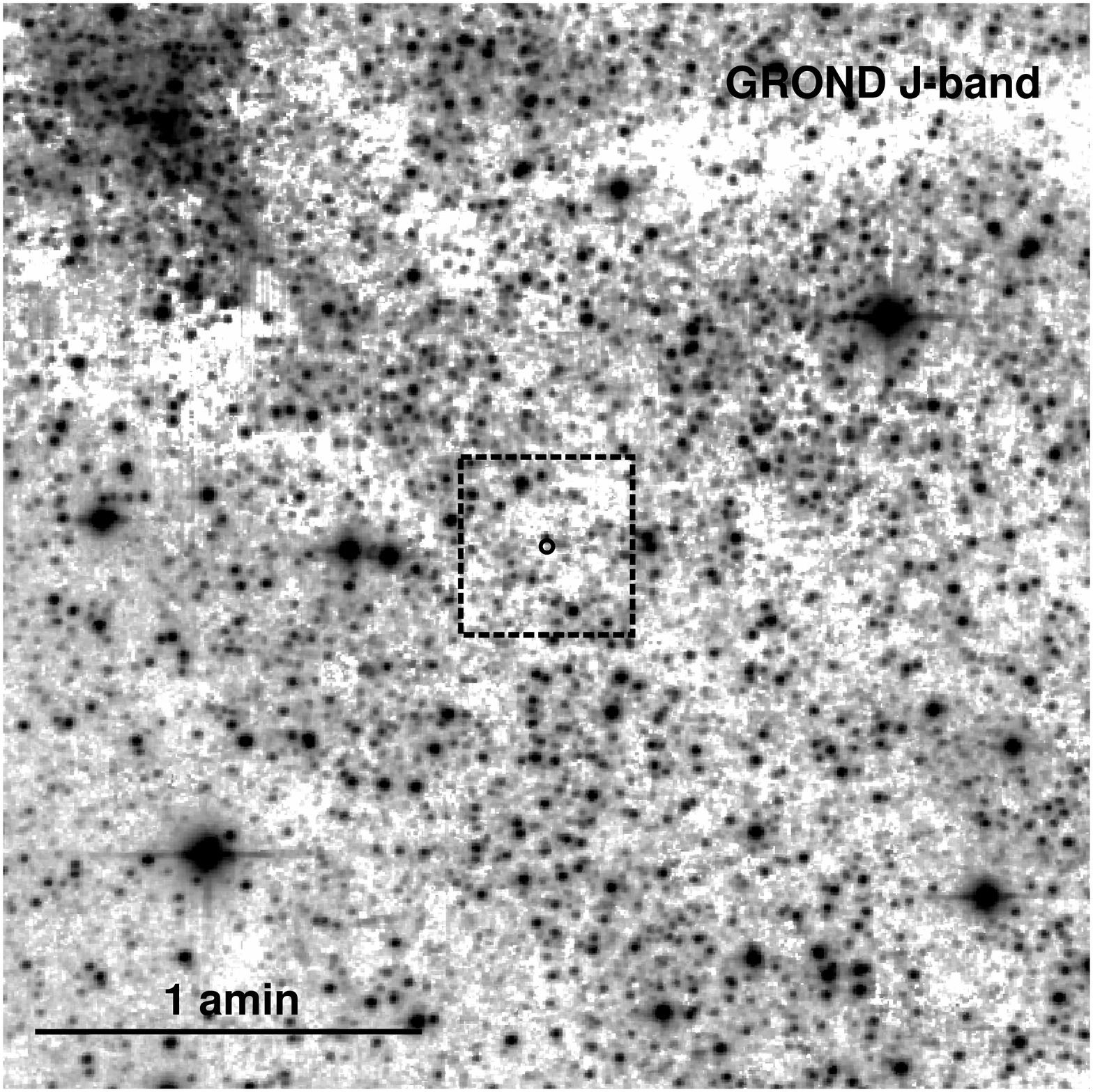}}
\fbox{\includegraphics[height=0.22\textwidth,angle=0]{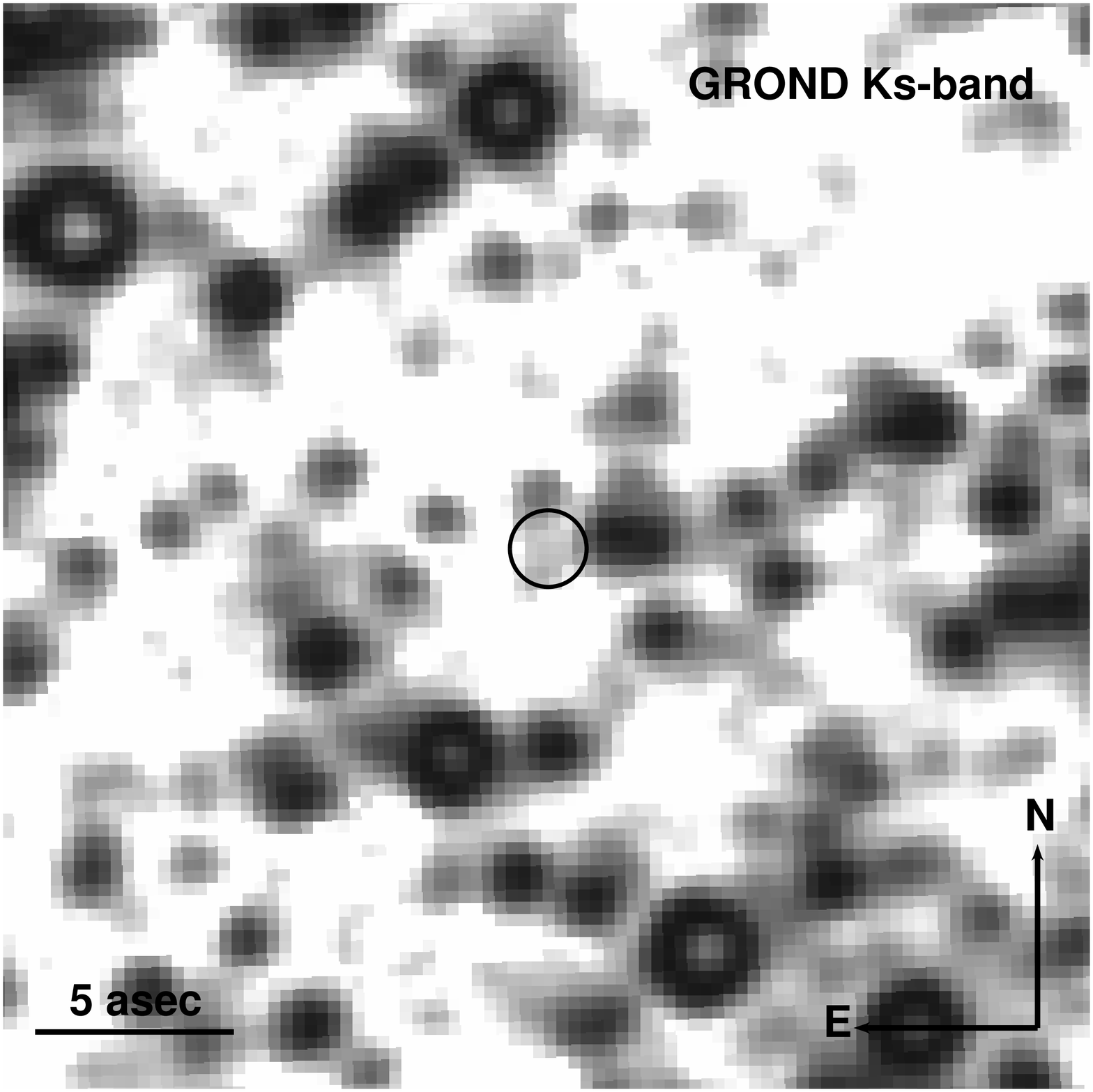}}
\caption{{\it Left:} GROND J-band image of the field of \axj. The 3$\sigma$ 
Chandra location uncertainty with radius of $0.96^{\prime\prime}$ 
is indicated by the black circle (Jin et al. 2017). The central star cluster 
surrounding \sgras\ appears in the top left corner of the image. 
{\it Right:} Zoom-in into the K$_s$-band image of the region shown 
by the dashed rectangle in the left image.}
\label{fig:grond_finder}
\end{figure}

Figure \ref{fig:grond_finder} (left) shows the J-band image of the region 
around \axj\ indicating the extreme crowding characteristic of the GC region. 
Figure~\ref{fig:grond_finder} (right) provides a zoom-in of the K$_s$-band image. 
We detect emission consistent with the 3$\sigma$ uncertainty of 
the \chandra\ position (see Jin et al. 2017a), but the available angular resolution 
does not allow us to distinguish between either emission from a single source 
or from the superposition of multiple objects. Therefore, we consider 
the photometry measurements of $J>19$, $H>15.5$ and $K_s>14.5$ 
(all in the Vega system), obtained with an aperture centred 
at the \chandra\ position, as upper limits on the observed magnitude 
of any counterpart to \axj. These magnitude upper limits correspond 
to flux densities of $J<4\times10^{-5}$\,Jy,  $H<7\times10^{-4}$\,Jy 
and $K<1\times10^{-3}$\,Jy.

Constraints on the donor star can be derived with an assumption 
on the distance to the source ($\approx8.3$\,kpc) and a good 
approximation of the Galactic foreground reddening. 
Using $A_{\rm J}\approx7.4$\,mag, $A_{\rm H}\approx4.2$\,mag, 
and $A_{\rm K}\approx2.4$\,mag (Fritz et al. 2011) the limits on 
the absolute magnitude of the companion star are estimated 
to be $M_{\rm J}>-3.0$\,mag, $M_{\rm H}>-3.3$\,mag, and 
$M_{\rm K}>-2.5$\,mag. 
This is in agreement with the classification of \axj\ as an LMXB 
and suggests the donor to belong to the stellar luminosity 
class III (giants) or fainter.

\subsection{\gmrt}
\label{gmrt}

The Giant Metrewave Radio Telescope (GMRT) is a multi-element 
aperture synthesis telescope consisting of 30 antennas, each 
with a diameter of 45 m, separated by a maximum baseline 
of 25 km (Swarup et al. 1997). We observed the GC region (centred at \sgras) 
at 1390 MHz on 2015 Aug 13.6 (UT) for 8 hours (DDTB178). 
Thanks to a simultaneous \chandra\ HETG observation, we can 
establish that \axj\ was in the soft state during the GMRT monitoring.
The primary beam FWHM of the GMRT is $24^\prime$ at 1.4 GHz, 
and hence \axj\ was well within 
the field of view of the observation. A total of 33.33 MHz bandwidth divided 
into 256 frequency channels was used with an integration time 
of 16.1 s. 3C~48 was used as the flux density calibrator, while the sources 
1822-096 and 1751-253 were used as phase calibrators.

The data were analysed using the Astronomical Image Processing 
System (AIPS). After flagging the original data set for non-working 
antennas and Radio Frequency Interference (RFI), data from a single 
frequency channel of the flux and phase calibrators were used 
to construct time-based amplitude and phase calibrations, while bandpass 
calibration was done with the flux calibrator. The bandwidth was divided 
into 32 channels of 1 MHz each before imaging, in order to minimize 
the effects of bandwidth smearing. A high resolution map of the region (imaged 
with baselines $1.5 k\lambda < b < 120 k\lambda$) was first constructed 
to resolve the diffuse emission in the region, followed by self-calibration. 
The clean components from this image (containing the resolved diffuse 
emission) were then subtracted from the original {\it ($u,v$)} data, which was used 
to construct a low resolution map of the region (baselines 
$0 k\lambda < b < 5 k\lambda$). In order to remove the confusing diffuse 
emission from the original map, the clean components from this low resolution 
image were then subtracted from the original {\it ($u,v$)} data. The final {\it ($u,v$)} data 
set was then used to construct a high resolution map (baselines 
$3 k\lambda < b < 120 k\lambda$) of the region with 3D-imaging over 31 facets, 
which was subsequently self-calibrated.

The high crowding and diffuse emission around the region limited 
the map RMS to $\approx$ 0.8 mJy/beam. No radio source was identified 
at the position of \axj\ with a 3 $\sigma$ upper limit of 2.4 mJy. 
The corresponding radio luminosity upper limit for the source 
is $2.7 \times 10^{29}$ ergs/s at 1.4 GHz.

\section{Accretion states and X-ray light curve}

\begin{figure}
\includegraphics[height=0.38\textwidth,angle=0]{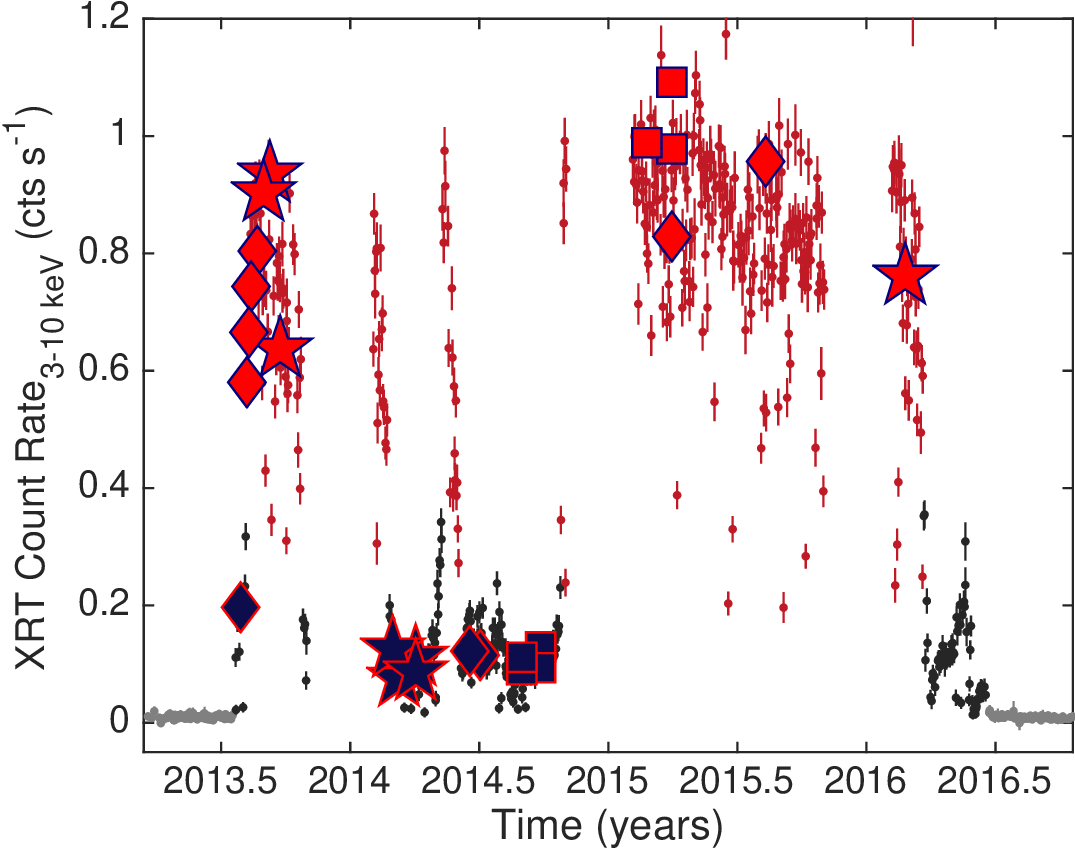}
\caption{\swift-XRT light curve of the latest outburst of \axj. 
The red, black and grey points indicate 
the soft, hard and quiescent source states, respectively (see text for more details). 
The stars, squares and diamonds indicate the time and equivalent \swift-XRT count 
rate for each of the \xmm, \nustar\ and simultaneous (\xmm+\nustar) observations, 
respectively. Red and black symbols manifest observations that caught \axj\ in the 
soft state and hard state, respectively. We do not remove \swift\ exposures (partially)
accumulated during eclipses. }
\label{cluce}
\end{figure}

\begin{figure}
\includegraphics[height=0.35\textwidth,angle=0]{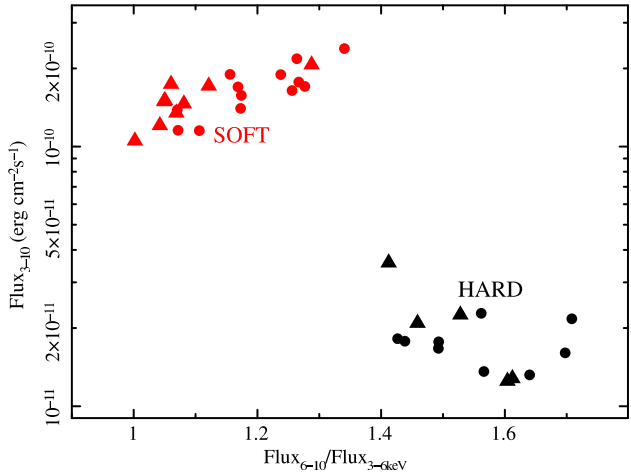}
\caption{Hardness intensity diagram (HID) of \axj\ derived 
from each \xmm\ and \nustar\ observation 
(they are represented with circles and triangles, respectively). 
Red and black symbols indicate soft and hard states, respectively. 
The two states are clearly separated. }
\label{HID}
\end{figure}

The dots in Fig. \ref{cluce} display the evolution of the 3-10 keV emission of \axj\ 
during its last outburst, as observed by \swift-XRT (see Degenaar et al. 2014 
and Ponti et al. 2015 for an historical light curve). The colour 
code of the circles provides an estimate of the source state\footnote{
We determine the state of \axj\ based on X-ray colours. However, we point out 
that the presence of eclipses and dips, that significantly modify the observed 
X-ray colours, make the determination of the source state very challenging
within the short (typically 0.5-1~ks) \swift\ exposures. We tentatively associate 
with a soft state the \swift\ observations with an average 3-10~keV flux 
in excess of $5\times10^{-10}$ erg cm$^{-2}$ s$^{-1}$ and hardness 
lower than 1.4 (e.g., see Fig. 3). However, to reduce the effect of dips 
and eclipses, the average is computed over 5 consecutive \swift\ observations. }
(e.g., grey, black, and red points correspond 
to the quiescence, hard and soft states, respectively). 
Although the determination of the spectral state based 
uniquely on \swift\ data is fairly uncertain (because of the large uncertainties 
associated with the X-ray colour within the short \swift-XRT exposures) we note 
that they, in all cases, agree with the states deduced from the higher quality 
\xmm\ and \nustar\ data (Fig. \ref{cluce} and \ref{HID}). 

The light curve of \axj\ exhibits many interesting behaviours. 
During the latest outburst, \axj\ went at least four times through the 
entire hysteresis loop (from a hard to soft and back to the hard state), 
before going back to quiescence. This behaviour is observed in several 
other NS LMXB and it appears starkly different from outbursts of BH LMXB, 
which typically perform the hysteresis loop only once 
(Mu\~{n}oz-Darias et al. 2014; Ponti et al. 2014). 

Similar to Ponti et al. (2015), we determine the state of the source 
on the basis of the X-ray colour. Figure \ref{HID} shows the hardness intensity 
diagram (HID), which is often used to determine the source state (Fender et al. 
2004; Belloni et al. 2011; Mu\~{n}oz-Darias et al. 2014). The hardness is 
defined here as the ratio between the observed fluxes in the 6-10 and 
3-6~keV bands. As it is rather widespread in Atoll sources, we note that during 
all \xmm\ and \nustar\ observations (which allow us to securely pin down 
the state of the source) the flux of \axj\ during the hard state is in all cases 
significantly lower than in the soft state (see Fig. \ref{cluce} and \ref{HID}). 

\section{New \xmm\ observations} 
\label{fitxmm}

\subsection{\xmm\ soft state} 
\label{xmmsoft}

\begin{figure}
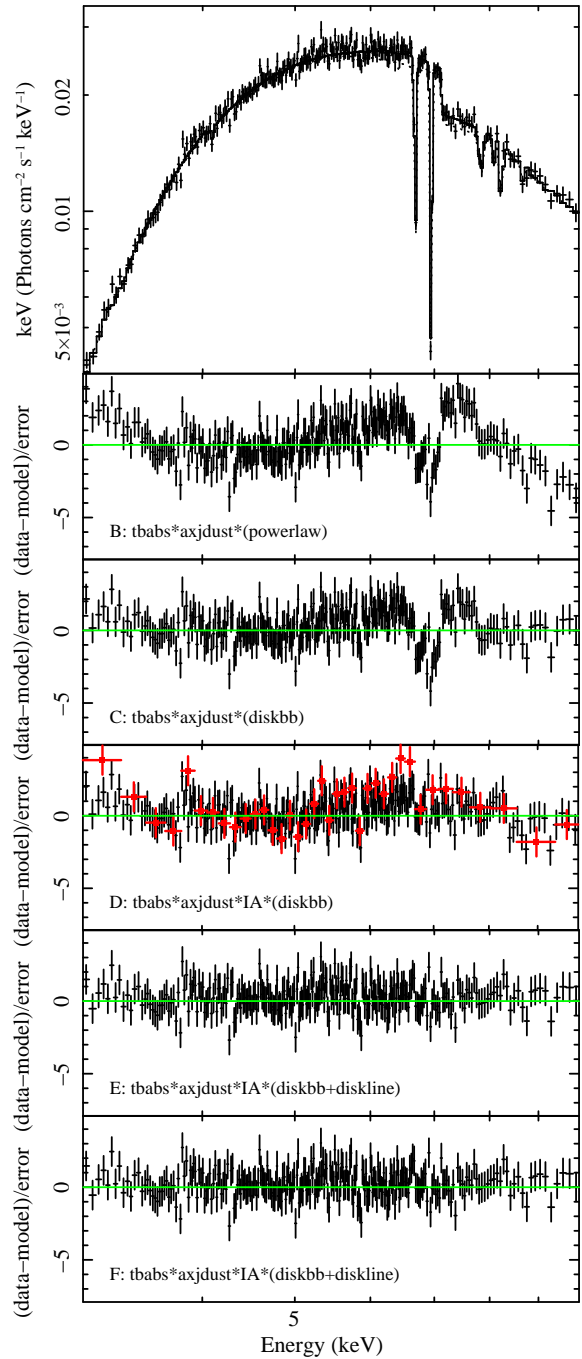

\includegraphics[height=0.45\textwidth,angle=-90]{XMMnewSoftTbabsCloudyDisklineDiskbbeuf.ps}
\vspace{-0.28cm}

\includegraphics[height=0.45\textwidth,angle=-90]{XMMnewSoftTbabsPLdel.ps}
\vspace{-0.49cm}

\includegraphics[height=0.45\textwidth,angle=-90]{XMMnewSoftTbabsDiskbbdel.ps}
\vspace{-0.49cm}

\includegraphics[height=0.45\textwidth,angle=-90]{XMMnewSoftTbabsCloudyDiskbb2datasetsdel.ps}
\vspace{-0.49cm}

\includegraphics[height=0.45\textwidth,angle=-90]{XMMnewSoftTbabsCloudyDisklineDiskbbdel.ps}
\vspace{-0.49cm}

\includegraphics[height=0.45\textwidth,angle=-90]{XMMnewSoftTbabsCloudyDisklineDiskbbdelFreez.ps}
\caption{\xmm\ combined spectrum of the three new soft state 
observations (Tab. \ref{data}). {\it Top panel:} Unfolded spectrum, 
fitted with the best fit model, composed by disk black body plus 
disk-line emission absorbed by neutral and ionised material 
({\sc tbabs*axjdust*IA*(diskbb+diskline)}). 
{\it Panel B:} Observed residuals, once the spectra are fitted with 
an absorbed power law model ({\sc tbabs*axjdust*(powerlaw)}). 
{\it Panel C:} As panel B, once the power-law is substituted with a disc 
black body model ({\sc tbabs*axjdust*(diskbb)}). Clear residuals 
are still present. 
{\it Panel D:} As panel C, once an ionised absorption model 
component is added to the fit ({\sc tbabs*axjdust*IA*(diskbb)}). 
The red points show the residuals, once the data are heavily 
rebinned. Small remaining residuals in the 5-8 keV band are still visible. 
{\it Panel E:} Observed residuals, once a broad Fe K$\alpha$ 
emission line is added to the best-fit ({\sc tbabs*axjdust*IA*(diskbb+diskline)}). 
{\it Panel F:} As panel E, once forcing the ionised and neutral absorption 
component to be constant over time, between the different soft state 
observations. }
\label{resXMM}
\end{figure}

The top panel of Fig. \ref{resXMM} shows the combined spectrum 
of the three new \xmm\ soft state observations (Tab. \ref{data}). 
We simultaneously fit the three spectra with the same model, 
however allowing the parameters to be different between
the three spectra as detailed below, and plot the combined 
data and residuals.  Consistent with what was observed 
in archival observations (Ponti et al. 2015), the X-ray emission 
is absorbed by a large column density of neutral (lowly ionised)
equivalent hydrogen also during the new \xmm\ observations 
(Fig. \ref{resXMM}).  

We start by simultaneously fitting these three new soft state \xmm\ spectra with 
a disc black body model, absorbed by neutral material and modified 
by dust scattering effects ({\sc tbabs*axjdust*diskbb} in \xspec; see 
Jin et al. 2017a,b for a full description of the {\sc axjdust} component). 
The free parameters in the fit of each spectrum 
are: the column density of neutral hydrogen absorption ($N_H$), 
the disk black 
body temperature ($kT_{DBB}$) and disc normalisation ($N_{DBB}$). 
The fit leaves highly significant residuals between 
$\sim6-8$~keV, clear signatures of the presence of an additional 
ionised absorption component (see panel C of 
Fig. \ref{resXMM}; $\chi^2_{DBB}=2162.5$ for 2037 degree of freedom; 
dof). 

Before investigating further the nature of the residuals in the 
$\sim6-8$~keV band, we tested the shape of the broad band spectrum. 
We first substituted the thermal disc black body component with 
a power law ({\sc tbabs*axjdust*powerlaw}) with $N_H$, power-law 
photon index $\Gamma$ and normalisation $N_{pl}$ as 
free parameters. We observed a significantly worse fit 
compared to the thermal disc black body model (see panel B of 
Fig. \ref{resXMM}; $\chi^2_{PL}=2525.5$ for 2037 dof). 
Therefore, we can exclude that the continuum during 
these soft state spectra can be reproduced by a power law. 
Finally, we exchanged the disk black body component 
with a black body one ({\sc tbabs*axjdust*bbody}) with $N_H$, black body 
temperature $kT_{BB}$ and black body radius $r_{BB}$ as free 
parameters. In this case, the fit is acceptable and comparable to 
the one employing the disc black body continuum ($\chi^2=2135.8$ 
for 2037 dof). 
We concluded that, consistently with the results obtained from archival 
data (Ponti et al. 2015), the new soft state observations are consistent 
with an absorbed thermal continuum model, while a power law is excluded. 
We also noted that the two thermal continuum models appear degenerate 
within the limited \xmm\ energy band. We, therefore, report in the rest of 
this section only the results obtained by fitting the spectra with the disc 
black body model. We also note that the inferred best fit inner disc 
radius results to be $r_{DBB}\sim3.5$~km smaller than the NS radius. 
This suggests that the underlying continuum is more complex than 
statistically required by the \xmm\ data alone. Indeed, multiple emission 
components considering either and both black body and disc black body 
components will be explored whenever simultaneous \nustar\ data will 
be available; Tab. 2 and 3).

To reduce the residuals observed in the $\sim6-8$~keV energy 
band, we added a self consistent ionised absorption component 
($IA$), accurate for the soft state SED of \axj\ (see Ponti et al. 
2015 for details)\footnote{The self consistent ionised absorption 
component has been created through the photo-ionisation code 
Cloudy C13.00 (Ferland et al. 2013). The model ingredients are: 
(1) the soft and hard spectral energy distributions as determined 
in Ponti et al. 2015; (2) constant electron density $n_e=10^{12}$ 
cm$^{-3}$; (3) ionisation parameter in the range 
$log(\xi/1$ erg cm s$^{-1})=23:24.5$; (4) intervening column density 
in the range $log(N_H/1$ cm$^{-2})=23.0:24.5$; (5) turbulent velocity 
$v_{turb}=500:1000$ km s$^{-1}$; (6) chemical abundances as in 
table 7.1 of Cloudy documentation. }. 
We leave as free parameters for each spectrum 
the plasma ionisation parameter ($log(\xi_{IA})$) and column density 
($log(N_{H_{IA}})$). We leave the absorber outflow velocity ($v_{out}$) 
free to vary, however we force it to be the same for all spectra. 
The addition of such component provides a highly significant improvement 
of the fit (see panel D of Fig. \ref{resXMM}; 
$\Delta\chi^2=163.6$) as well as a satisfactory description 
of the data ($\chi^2=1998.9$ for 2030 dof). 

By heavily rebinning the data (see red points in panel D of 
Fig. \ref{resXMM}), we noted some remaining broad residuals in the 
$\sim6-7.5$~keV band. This excess emission (panel D of Fig. \ref{resXMM}) 
is observed to be placed red-ward of the absorption features
and it appears to be reminiscent of P-Cygni profiles observed in some BH 
systems and due to the combination of emission and absorption 
from an outflowing plasma (King et al. 2015; Miller et al. 2015; 
Munoz-Darias et al. 2016; 2017). 
Therefore, we model this emission component with the addition 
of a Gaussian line ({\sc tbabs*IA*axjdust*(diskbb+gaussian})). The line 
energy and width are free to vary, but tied between the three spectra. 
The addition of the Gaussian line provides a significant improvement 
of the fit ($\Delta\chi^2=79.0$ for 6 more free parameters; 
$\chi^2=1919.9$ for 2024 dof; F-test probability $<10^{-6}$). 
The best-fit energy, width and equivalent width of the line are: 
$E=6.43\pm0.25$~keV, $\sigma=0.85^{+0.23}_{-0.17}$~keV 
and $EW\sim140-200$~eV. The line is too broad to be the 
redshifted emitted component of the same outflowing plasma 
producing the absorption features. Indeed, the observed line 
broadening ($\sigma\sim0.7-1.0$~keV) would require bulk 
outflows of $v_{out}\sim0.1$~c. We can rule out that the 
ionised absorbing plasma is outflowing at such a large 
speed (e.g., $v_{out}<2000$~km~s$^{-1}$). 

Instead, such residual emission might be due to a broad 
emission line, reflected off the accretion disc. 
Indeed, in such a scenario, the disc line would be expected to 
appear highly broadened as a consequence of the high disc inclination. 
We, therefore, added a diskline component to the fit 
({\sc tbabs*IA*axjdust*(diskbb+diskline)}), assuming the line energy, inner 
and outer disc radii to have the values: $E=6.4$~keV, $r_{in}=6$ and 
$r_{out}=10^3$~$r_g$ ($r_g=GM/c^2$, where G is the gravitational 
constant, $M$ is the mass of the compact object and $c$ the speed of light). 
We also assumed the same emissivity profile 
and disc inclination, for all spectra. We observed a highly significant 
improvement of the fit (compared to the same model without 
the disk line) adding such a disk line component to the 
model (see panel E of Fig. \ref{resXMM}; $\Delta\chi^2=78.1$ 
for 6 more free parameters; $\chi^2=1920.8$ for 2024 dof). 
The best-fit disc inclination is $i={70_{-15}^{+7}}^\circ$ 
(on the lower range expected for eclipsing systems, however consistent 
with the eclipsing nature of the source) 
and the emissivity index is scaling with disc radius as $r^{-2.4\pm0.1}$. 
The line is observed to have an equivalent width in the range 
$EW=120-200$~eV, therefore consistent with reflection off a standard 
accretion disc (Matt et al. 1993). 
This fit provides an acceptable description of the data 
(Tab. \ref{fitxmmsoft} and Fig. \ref{resXMM}). 

Alternatively, the residuals observed in the Fe~K band might be 
associated with an improper characterisation of the Fe~K edge 
imprinted by the neutral absorption (e.g., due to abundances different 
from Solar or due to depletion; Ponti et al. 2016). 
To test this latter possibility, we fitted the data without a broad 
iron line component, but leaving the iron abundance of the neutral absorber 
free to vary. We observed only a marginal improvement of the fit 
($\Delta\chi^2=4.9$ for 1 more dof), with a best-fit iron abundance of 
$Fe=1.13^{+0.16}_{-0.09}$~Solar. Therefore, we do not consider 
this alternative possibility any further. 

We observed that all parameters of both the highly and minimally ionised 
absorption\footnote{Dips have been filtered out in the analysis 
presented in this paper (see \S \ref{xmmDR}). } components are 
consistent with being 
constant among the new observations as well as consistent with previous 
observations (see Tab. \ref{fitxmmsoft} and Ponti et al. 2015)\footnote{The 
apparently different best fit column density of neutral absorption obtained 
fitting the new data, compared to archival observations, is due 
to the different absorption models (e.g., here the {\sc tbabs} instead than 
the {\sc phabs} model is used), to the assumed abundances and cross 
sections as well as an improved treatment of the effects of dust scattering 
(see Ponti et al. 2017 for a more detailed description). }. We, therefore, 
fit all the new soft state spectra forcing the absorption to remain 
constant over time. As a result the best fit parameters are: 
$N_H=29.0\pm0.4\times10^{22}$~cm$^{-2}$, 
$log(\xi_{IA})=3.92\pm0.16$, $log(N_{H_{IA}})=23.3\pm0.3$ 
and $v_{turb}=700$~km~s$^{-1}$ and no significant variation of the fit is 
observed (see panel F of Fig. \ref{resXMM}; $\chi^2=1926.0$ 
for $2030$ dof, $\Delta\chi^2=-5.2$ for the elimination of 6 dof; 
Tab. \ref{fitxmmsoft}). 

Hereinafter we will refer to the combination of absorption and 
scattering effects {\sc tbabs*IA*axjdust} observed during the soft state 
as {\sc softabs}. 
\begin{table*}
\begin{center}
\small
\begin{tabular}{ l c c c c c r r c c c c c c c }
\hline
\hline
\multicolumn{8}{c}{\bf \xmm\ soft state} \\
\hline
\hline
\multicolumn{8}{l}{tbabs*axjdust*IA*(diskbb+diskline)} \\
{\sc OBSID}  & $N_{H}$           & log($\xi_{IA}$)                 & $log(N_{H_{IA}})$   &$N_{dl}$    & $kT$                & $N_{DBB}$    & $\chi^2/dof$\\
                    &($10^{22}$~cm$^{-2}$)&                      &                                 & $\dag$     &(keV)                  &  $\flat$ & \\
\hline
0743630901  & $29.2\pm1.1$ & $4.0^{+0.7}_{-0.1}$ & $23.2^{+0.6}_{-0.4}$&$12\pm5$& $2.37\pm0.10$ & $1.1\pm0.2$ & $642.8/676$ \\ 
0743630801  & $28.8\pm0.6$ & $3.9\pm0.2$          & $23.4\pm0.4$           &$10\pm4$& $2.34\pm0.06$ & $1.1\pm0.1$ & $647.9/676$ \\ 
0743630601  & $29.1\pm0.6$ & $4.0\pm0.2$            & $23.4^{+0.3}_{-0.2}$& $6\pm2$ & $2.14\pm0.05$ & $1.5\pm0.1$ & $630.0/676$ \\ 

                      & $29.6\pm0.4$ & $4.4\pm0.1$ & $23.9\pm0.3$ \\
\hline
\hline
\multicolumn{8}{c}{\bf \xmm\ hard state} \\
\hline
\hline
\multicolumn{6}{l}{tbabs*axjdust*power-law} \\
{\sc OBSID}  & $N_{H}$                     &&&& $\Gamma$ & $N_{pl}$ & $\chi^2/dof$\\
                     &($10^{22}$~cm$^{-2}$)&&&&                 & \ddag \\
\hline
0743630501 & $28.9\pm1.1$ &&&& $1.98\pm0.07$ & $1.8\pm0.3$ & $609.9/578$ \\ 
0743630401 & $29.6\pm1.5$ &&&& $2.06\pm0.09$ & $2.1\pm0.4$ & $624.2/578$ \\ 
0743630301 & $27.1\pm1.4$ &&&& $1.91\pm0.09$ & $1.5\pm0.3$ & $596.4/578$ \\ 
0743630201 & $29.3\pm1.3$ &&&& $2.12\pm0.08$ & $2.4\pm0.4$ & $545.0/578$ \\ 
0723410501 & $29.8\pm1.3$ &&&& $1.95\pm0.08$ & $1.3\pm0.2$ & $576.7/578$ \\ 
0723410401 & $27.9\pm1.3$ &&&& $1.97\pm0.08$ & $1.3\pm0.2$ & $595.3/578$ \\ 
0723410301 & $29.4\pm1.0$ &&&& $1.98\pm0.06$ & $2.3\pm0.3$ & $614.4/578$ \\ 
\hline
\hline
\end{tabular}
\caption{Best fit parameters of the new \xmm\ soft and hard state 
spectra, once fitted with single components absorbed models. 
The soft state spectra are best fit with absorbed thermal models with 
the additional absorption by ionised material, while the hard state 
spectra are best fit by absorbed power law emission. 
The best fit emissivity profile of the disk line is $r^{-2.4\pm0.1}$ 
reflected off a disc with an inclination of $i={70_{-15}^{+7}}^\circ$. 
During the soft state, the neutral and ionised absorption are 
consistent with being constant over time. However, the ionised 
absorption component disappears during the hard state. 
This is consistent with the results from archival 
observations. $\dag$ in units of $10^{-4}$ photons cm$^{-2}$ s$^{-1}$; 
\ddag~in units of $10^{-2}$ ph keV$^{-1}$ cm$^{-2}$ s$^{-1}$ at 1 keV; 
$\flat$~normalisation $N_{DBB}=R^2_{DBB}/D^2_{10}cos(\theta)$, 
where $R_{DBB}$ is the apparent inner disc radius in km, $D_{10}$ is 
the distance to the source in units of 10~kpc and $\theta$ the angle 
of the disc ($\theta=0$ is face on). }
\label{fitxmmsoft}
\end{center}
\end{table*} 

\subsection{\xmm\ hard state}
\label{xmmhard}

\begin{figure}
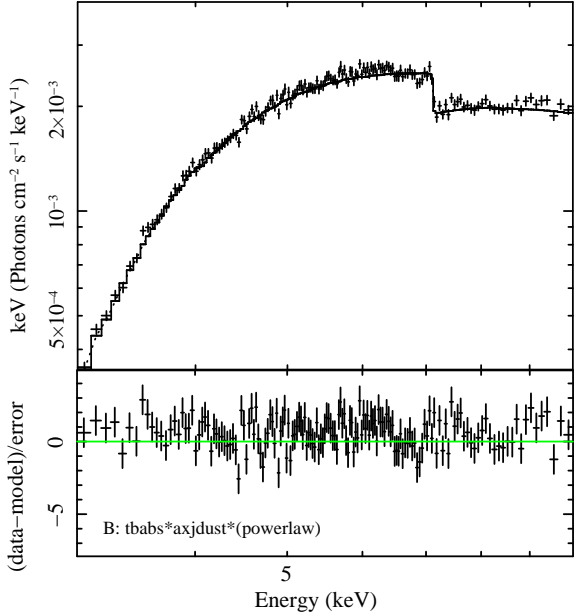

\includegraphics[height=0.45\textwidth,angle=-90]{XMMnewHardTbabsPLGaus2ld.ps}
\vspace{-0.28cm}

\includegraphics[height=0.45\textwidth,angle=-90]{XMMnewHardTbabsPLGaus2del.ps}
\caption{\xmm\ combined spectrum of the seven new hard state 
observations (Tab. \ref{data}). {\it Top panel:} Unfolded spectrum, 
fitted with the best fit model, composed by a power law absorbed by 
neutral material ({\sc tbabs*axjdust*(powerlaw)}). 
{\it Bottom panel:} Residuals, after fitting the data with the best fit model
({\sc tbabs*axjdust*(powerlaw)}). }
\label{resXMMhard}
\end{figure}
We performed a simultaneous fit of the seven new hard state \xmm\ spectra.
The absorbed power-law model provides a superior fit ($\chi^2_{PL}=4155.7$ 
for 4046 dof), compared to either an absorbed black body ($\chi^2_{BB}=4541.7$ 
for 4046 dof) or an absorbed disk black body ($\chi^2_{BB}=4259.9$ for 4046 dof). 

Despite no clear residuals are observed in the Fe~K band (see Fig. 
\ref{resXMMhard}), we add the same ionised 
absorption component, with the best fit parameters observed during the soft state. 
This results in a significant worsening of the fit ($\Delta\chi^2=-17.6$ for 
the same dof). We further check for the presence of ionised Fe~K lines. 
No narrow absorption lines are detected at more than $\sim95$~\% significance 
(a narrow Fe~{\sc xxvi} absorption line is detected at $\sim93$~\% and 
$\sim90$~\% significance, during observation 0743630201 and during 
0723410401, with an equivalent width EW$\simeq-16$ and
EW$\simeq-13$~eV, respectively, while we note that in the soft state 
such lines are typically observed with an equivalent width of 
EW$\simeq-25$~eV, EW$\simeq-30$~eV). 

From the spectra of the new hard state observations we can derive 
upper limits to the presence of narrow Fe~{\sc xxv} and Fe~{\sc xxvi} 
K$\alpha$ absorption lines between $EW_{Fe~{\sc xxv}}>-15$  
and $EW_{Fe~{\sc xxv}}>-30$~eV and $EW_{Fe~{\sc xxvi}}>-15$ 
and $EW_{Fe~{\sc xxvi}}>-35$~eV, respectively). 
This rules out the presence of the same ionised absorption 
component in the hard state, ubiquitously observed 
during the soft state. However, because of the shorter exposure, 
we note that these upper limits are less stringent than those 
obtained from the longer exposure of the 2014-04-03 observation 
(see Ponti et al. 2015). Therefore we can not exclude the presence 
of absorption lines with EW of -5:-15~eV in these new observations. 

We also tested that the addition of either an Fe K$\alpha$ disk-line or 
Gaussian emission line does not improve significantly the fit ($\chi^2=4153.9$ 
for 4038 dof), with upper limits to its equivalent width (assuming 
$\sigma=0.1$~keV) of the order of $EW<30$~eV. 

As observed in previous hard state observations, the neutral absorption 
column density is consistent with being constant between the various hard 
state observations (Tab. \ref{fitxmmsoft}; $\Delta\chi^2=-8.9$ for the 
elimination of 6 dof). 
We also note that the best fit equivalent neutral hydrogen column density
is $N_H=(28.9\pm0.5)\times10^{22}$~cm$^{-2}$, consistent with the value observed 
during the soft state, when fitted with the disk black body component (see 
Tab. \ref{fitxmmsoft} and Ponti et al. 2015).

Hereinafter we will refer to the absorption and scattering components 
{\sc tbabs*axjdust} observed during the hard state as {\sc hardabs}. 

\section{Simultaneous \xmm\ and \nustar\ observations}
\label{fitxmmNu}

Seven \xmm\ observations are simultaneous to five \nustar\ ones 
(see Tab. \ref{data} and \ref{TabNu}). To achieve the best possible 
constraints on the evolution of the X-ray SED, we performed a combined 
fit of these simultaneous \xmm\ and \nustar\ data. The joint \xmm\ and 
\nustar\ fit revealed a discrepancy at low energy. 
Full details and discussion about this discrepancy are provided 
in appendix \ref{NDR}. 

\subsection{\nustar+\xmm\ soft state: longest observation on 2015-02-26} 

\begin{table*}
\small
\begin{tabular}{ l l c c c c c c c c c c c c c c c c}
\hline
\hline
\multicolumn{10}{c}{\bf Simultaneous \xmm+\nustar\ soft state (2015-02-25 and -26)} \\
\hline
Model          &{\sc dbb-bb-dl}&{\sc simpdbb-dl}&{\sc dbb-nth-dl}&{\sc bb-nth-dl}&{\sc dbb-bb-rr}&{\sc dbb-nth-rr}&{\sc bb-nth-rr}&{\sc dbb-bb-nth-rr} & \\
$N_{H}$         &$30.9\pm0.6$   &$30.2\pm0.6$   &$31.1\pm0.6$    &$28.2\pm0.8$    &$31.6\pm0.6$  &$31.4\pm0.9$  &$27.9\pm0.5$&$32.2\pm0.9$ \\
$log(\xi_{IA})$ & $4.0\pm0.2$   & $4.2\pm0.2$   & $4.0\pm0.2$    & $4.0\pm0.2$    &$4.1\pm0.2$   &$4.1\pm0.2$   &$4.2\pm0.2$  &$4.1\pm0.2$  \\
$log(N_{H_{IA}})$& $23.5\pm0.3$  & $23.8\pm0.3$  & $23.5\pm0.2$   & $23.5\pm0.3$    &$23.7\pm0.2$ &$23.6\pm0.2$  &$23.6\pm0.3$ &$23.5\pm0.2$ \\
$kT_{DBB}$      & $1.77\pm0.05$ & $1.70\pm0.09$ & $1.63\pm0.09$  &                 &$1.55\pm0.11$&$1.1\pm0.1$   &             &$1.1\pm0.1$  \\
$N_{DBB}$       & $3.0\pm0.4$   & $3.3\pm0.6$   & $2.8\pm0.9$    &                 &$5.1\pm1.7$  &$7.0\pm4.2$   &             &$17\pm6$     \\
$kT_{BB}$       & $3.0\pm0.1$   &               &                &$1.06\pm0.08$    &$3.1\pm0.2$  &              &$1.01\pm0.09$&$2.3\pm0.2$  \\
$N_{BB}$        &$0.12\pm0.03$  &               &                &$0.015\pm0.011$  &$0.08\pm0.04$&              &$7\pm3$      &$0.5\pm0.2$ \\
$f_{sc}$        &               &$0.63\pm0.02$  &                &                 &             &              &             &             \\
$\Gamma$       &               &$4.6\pm0.03$   &$2.2\pm0.4$     & $2.6\pm0.6$     &             &$2.1\pm0.3$   &$2.1\pm0.2$  &$1.1^{+1.1}_{-0.2}$\\
$kT_{e}$        &              &               &$3.6^{+1.2}_{-0.4}$&$3.8^{+0.2}_{-0.6}$  &            &$3.3\pm0.3$   &$3.3\pm0.3$  &$4.6^{+10.3}_{-2.4}$\\
$N_{nth}$       &               &               &$2.1^{+3.8}_{-2.0}$& $1.1\pm0.7$      &            &$3.7\pm2.1$   &$0.8\pm0.2$  &$0.01\pm0.01$\\
$N_{dl}$        &$8.3\pm1.4$    & $7.4\pm1.3$   & $8.7\pm1.2$    & $7.9\pm1.5$      &            &              &             &             \\
$N_{ref}$       &               &               &                &                  &$36\pm9$&$55\pm11$    &$57\pm8$     &$82\pm16$     \\
$c_{NuA}$       &$1.08\pm0.02$  & $1.08\pm0.02$ & $1.08\pm0.01$  &$1.08\pm0.01$     &$1.08\pm0.01$&$1.08\pm0.01$&$1.08\pm0.01$&$1.08\pm0.01$\\
$c_{NuB}$       & $1.12\pm0.02$ & $1.12\pm0.02$ & $1.12\pm0.01$  &$1.12\pm0.01$     &$1.12\pm0.01$&$1.12\pm0.01$&$1.12\pm0.01$&$1.12\pm0.01$\\
$\chi^2/dof$   & 1348.5/1360   & 1397.2/1360   & 1346.9/1359    & 1344.1/1359      &1318.3/1360  &1306.9/1359  &1328.2/1359  &1288.5/1357  \\
\hline
\hline
\end{tabular} 
\caption{Best fit model of the longest simultaneous soft state \xmm\ 
and \nustar\ observations of \axj\ (2015-02-25 and 26). 
The different lines in the Table report: 
the model name, the equivalent hydrogen column density 
of neutral absorption (in units of $10^{22}$ atoms cm$^{-2}$), 
the logarithm of the ionisation parameter ($log(\xi_{IA})$) and column 
density ($log(N_{H_{IA}})$) of the ionised absorber, the temperature 
($kT_{DBB}$, in keV) and normalisation ($N_{DBB}$) of the disk black body 
and black body components ($kT_{BB}$ in keV and $N_{BB}$) 
(the disc black body normalisation is: 
$N_{DBB}=R^2_{DBB}/D^2_{10}cos(\theta)$, where $R_{DBB}$ is the 
apparent inner disc radius in km, $D_{10}$ is the distance to 
the source in units of 10~kpc and $\theta$ the angle of the disc; 
the black body normalisation is: $N_{BB}=r^2_{BB}/D^2_{10}$, 
where $r_{BB}$ is the black body radius in km and $D_{10}$ is 
the distance to the source in units of 10~kpc),
the fraction of the Comptonised component ($f_{sc}$), the asymptotic photon 
index of the power law ($\Gamma$), electron temperature ($kT_e$ in keV) 
and normalisation of the Comptonised component ($N_{nth}$ in $10^{-2}$ 
units), the normalisation of the disk-line component ($N_{dl}$ in units of 
$10^{-4}$ photons cm$^{-2}$ s$^{-1}$), the normalisation of the 
ionised-relativistic reflection component ($N_{ref}$ in units of $10^{-26}$ 
photons cm$^{-2}$ s$^{-1}$), the cross-normalisation constants 
($c_{NuA}$ and $c_{NuB}$ for \nustar\ module A and B, respectively) 
and statistic ($\chi^2$ and dof). }
\label{xmmNuSoft}
\end{table*} 

We started by fitting the longest simultaneous \xmm\ plus \nustar\ 
observation that caught \axj\ in the soft state, fitting the low energy 
excess with the modified absorption model (\S \ref{crosscal}). 
We initially employed the single emission component model used for the \xmm\ 
data (\S~\ref{xmmsoft}). Therefore, we fitted the spectra with an absorbed 
disk black body plus disk line model ({\sc softabs*(diskbb+diskline)}). 
This fit left large residuals (see 
panel "single" of Fig. \ref{xmmNuSTARsoft}) at high 
energies, producing an unacceptable fit ($\chi^2=2225.6$ for 1362 dof). 
These residuals demonstrated the need for a second broad band spectral 
emission component. 

\subsubsection{Two thermal emission components plus disk-line}
\label{2thdiskline}

We then added to the model an extra black body component 
({\sc softabs*(diskbb+bbody+diskline)}, dubbed {\sc dbb-bb-dl}), producing 
a significant improvement of the fit ($\Delta\chi^2=881.7$ for 2 new 
free parameters; 
$\chi^2=1343.9$ for 1360 dof).  With this model, the spectra are best fit 
by a prominent and hot ($kT_{BB}\sim1.19\pm0.03$~keV) black body component 
produced from a surface with a reasonable (however small) radius of 
$r_{BB}=3.0\pm0.2$~km. However, the high energy emission is produced 
by a very hot disc black body emission ($kT_{DBB}=3.4\pm0.1$~keV) with 
an inner radius of the accretion disc ($r_{DBB}\sim1$~km) significantly smaller 
than the NS radius. For this reason, this result is unphysical and it will not 
be discussed any longer. 

\begin{figure}
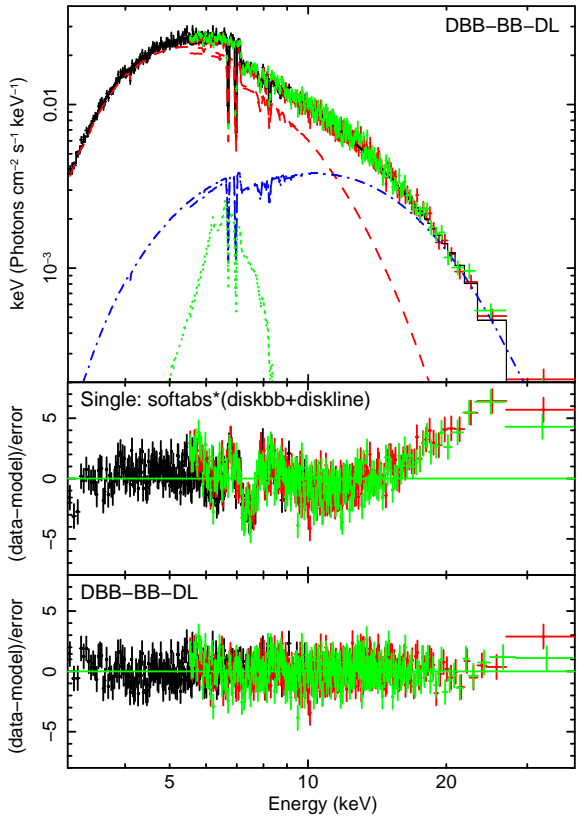

\includegraphics[height=0.45\textwidth,angle=-90]{DBB-BB-DLeuf.ps}
\vspace{-0.1cm}

\includegraphics[height=0.45\textwidth,angle=-90]{SingleDel.ps}
\vspace{-0.10cm}

\includegraphics[height=0.45\textwidth,angle=-90]{DBB-BB-DLdel.ps}
\caption{\xmm+\nustar\ spectra of the longest simultaneous 
soft state observation obtained on (2015-02-25 and 26). 
{\it Upper panel:} Broad band fit performed with the double 
thermal plus diskline model ({\sc dbb-bb-dl}). 
The black, red and green data show the \xmm, \nustar\ A and B
modules, respectively. The dashed red, dash-dotted blue, 
dotted green and solid black lines show the emission from 
the disk black body, black body, disk-line components and total 
emission, respectively. 
{\it Panel Single:} Residuals after fitting the spectra with the single 
thermal continuum model ($softabs*(diskbb+diskline)$) best fitting 
the soft state \xmm\ spectra. Clear residuals are left at high energy,  
implying the need for a second continuum component. 
{\it Lower panel:} Residuals left by the fit with an absorbed double 
thermal component model ({\sc dbb-bb-dl}) imposing a hotter 
disk black body component than the black body one. 
The model provide an acceptable description of the data. }
\label{xmmNuSTARsoft}
\end{figure}

We then fitted the spectrum with the same model, but imposing the 
temperature of the disk black body emission to be smaller than 
the one of the black body (Fig. \ref{xmmNuSTARsoft}; Tab. \ref{xmmNuSoft}). 
With this set-up, the spectra were fit by a warm ($kT_{DBB}=1.77\pm0.05$~keV) 
disc black body component that reproduced most of the observed X-ray emission. 
In particular, the best fit inner disc radius is several times larger 
than in the previous case ($r_{DBB}=5.7\pm0.4$~km), however still 
rather small. This time, the black body components appeared 
hotter ($kT_{BB}=3.0\pm0.1$~keV) and smaller ($r_{BB}\sim0.28\pm0.04$~km). 
Therefore, despite this set up provides 
a slightly worse fit ($\Delta\chi^2=-4.6$) compared to the previous set up, 
we preferred it on physical grounds. 

\subsubsection{Thermal plus Comptonisation and disk-line components}
\label{2thbbdl}

Alternatively, the high energy part of the emission of \axj\ could be 
produced by Comptonisation. 

\subsubsection*{High energy cut off in the X-ray band ({\sc simpl})}

To start, we parametrised the Comptonisation component by 
substituting the hot black body emission with a Comptonised radiation 
model which assumes that the high energy cut off is at very high energy
(the {\sc simpl} convolution model; Steiner et al. 2009). Therefore, we fitted the 
data with the model {\sc simpdbb-dl}: {\sc softabs*(simpl(diskbb)+diskline)}. 
This model provided a significantly worse fit compared to the previous 
models ($\chi^2=1397.2$ for 1360 dof; Tab. \ref{xmmNuSoft}). 

We note that the photon index of the Comptonised component yields 
an excessively steep value of $\Gamma=4.6\pm0.1$. 
Such steep spectra are typically signalling that either the Comptonised 
component has a thermal origin, or it indicates the presence of a high 
energy cut off in the power law shape. To test this latter possibility, 
we re-fitted the spectrum assuming a photon index of $\Gamma=2$ for 
the Comptonised component and we added to the model an exponential 
cut off\footnote{We assumed the folded energy to be equal to the cut off 
energy.} ({\sc highecut} in \xspec; {\sc softabs*simpl(diskbb+diskline)highecut}).
We obtained a comparable fit ($\Delta\chi^2=5.3$ for the same dof) for 
a cut off energy of $E_c=6.7\pm0.1$~keV. This indicates that if 
the high energy emission is due to a Comptonisation component, 
then the Comptonising electrons must have a temperature of a few keV, 
therefore about two orders of magnitude lower than what is 
observed in BH binaries and NS during the hard state (see \S 5.3). 

\subsubsection*{Comptonisation with high energy cut off ({\sc Nthcomp})}

We then applied a more sophisticated Comptonisation model that 
self consistently reproduces both the power law shape and the high 
energy cut off ({\sc nthcomp}; Zdziarski et al. 1996; Zycki et al. 1999). 
Therefore, we applied the model {\sc softabs*(diskbb+nthComp+diskline)}, 
that we call {\sc dbb-nth-dl}. 
We assumed the temperature of the seed Comptonised photons to be 
equal to the temperature of the disk black body component. 

This provides a good fit ($\chi^2_{DBB}=1346.9$ for 1359 dof; see 
Tab. \ref{xmmNuSoft}), slightly better than the double thermal 
emission model. In particular, we now observe that the asymptotic power law 
photon index is steep ($\Gamma\sim2.2$), however physically acceptable 
and within the range of values observed during the soft state in other NS 
and BH systems. 
Moreover, the temperature of the Comptonising electron is constrained to be: 
$kT_e\sim3-5$~keV, therefore producing a high energy cut off in 
the X-ray band, as also suggested by the fit with the {\sc simpl} component.
On the other hand, we note that the derived inner disc radius appears to be 
too small ($r_{DBB}=5.5\pm0.8$~km), even considering the large uncertainties 
associated with the derivation of this parameter, unless \axj\ is beyond 16~kpc 
from us. 

We also consider the alternative scenario where the thermal emission is 
produced by black body emission and not by the disk black body. We call 
this model {\sc bb-nth-dl}, {\sc softabs*(bbody+nthComp+diskline)}. 
We note that this model can also reproduce 
the data ($\chi^2_{BB}=1344.1$ for 1359 dof; Tab. \ref{xmmNuSoft}).

\subsubsection{Two thermal components plus relativistic ionised reflection}
\label{2thbbrefl}

The simultaneous \xmm+\nustar\ spectra are consistent with the 
presence of a broad Fe K$\alpha$ emission line during this soft state 
observation. The Fe K$\alpha$ line is often the most prominent feature 
of a reflection component (Fabian et al. 2000; 2009; Nandra et al. 2007; 
Ponti et al. 2006; 2010; Plant et al. 2014). 
We, therefore, tested for the presence of such a reflection component 
by substituting the {\sc diskline} with a {\sc bbrefl} model. 
The {\sc bbrefl} model reproduces a self consistent ionised reflection 
spectrum obtained by illuminating a constant density slab 
with a black body spectrum (Ballantyne 2004). 
We convolved the {\sc bbrefl} ionised reflection component 
with the {\sc kdblur} kernel, that is mimicking the relativistic effects 
on the shape of the reflection component off an accretion disc around 
a compact source (Laor 1991). 
We assumed a disc inclination, inner and outer radii of $80^\circ$, 
$6 r_g$ and $400 r_g$, respectively. We also fixed the value of 
the emissivity index to $\alpha=2.4$, such as derived by the fit with 
the disk line component, the Iron abundance to Solar and ionisation 
parameter to $log(\xi)=1$. 

\begin{figure}
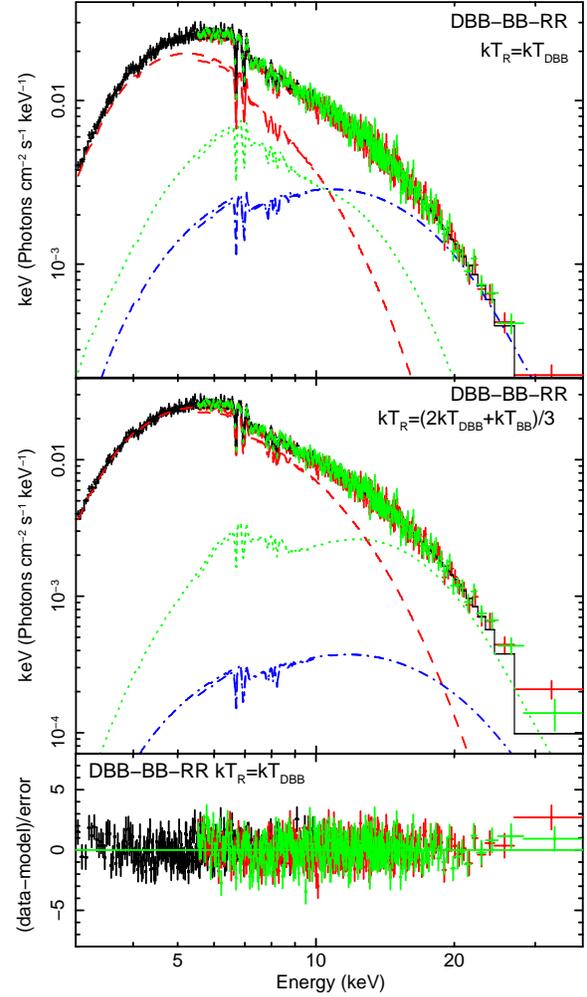

\includegraphics[height=0.45\textwidth,angle=-90]{DBB-BB-RReuf.ps}
\vspace{-0.1cm}

\includegraphics[height=0.45\textwidth,angle=-90]{DBB-BB-RRkT=2kTDBBkTBBeuf.ps}
\vspace{-0.1cm}

\includegraphics[height=0.45\textwidth,angle=-90]{DBB-BB-RRdel.ps}
\caption{\xmm+\nustar\ spectra of the longest simultaneous 
soft state observation, same as Fig. \ref{xmmNuSTARsoft}. 
{\it Upper panel:} Best fit broad band fit performed with the double 
thermal plus relativistic ionised reflection model ({\sc dbb-bb-rr}). 
The dashed red, dash-dotted blue, dotted green and solid black lines 
show the emission from the disk black body, black body, relativistically 
ionised reflection components and total emission, respectively. 
The reflection spectrum is produced by illuminating the disc with a thermal 
emission with $kT_R=kT_{DBB}$. 
{\it Middle panel:} Same as upper panel, but assuming that the 
disc is irradiated by a hotter thermal component 
($kT_R=(2\times kT_{DBB}+kT_{BB})/3$). 
{\it Lower panel:} Residual left once the data are fitted with the model 
shown in the upper panel. }
\label{xmmNuSTARsoft2}
\end{figure}

We fitted the double thermal plus relativistic ionised 
reflection model to the data: {\sc softabs*(diskbb+bbody+kdblur(bbrefl))}.
We call this model, {\sc dbb-bb-rr}. 
Because the disc black body component is dominating the source 
emission up to $\sim15$~keV, we imposed that the temperature of 
the illuminating black body emission is equal to the temperature of 
the disk black body component. This model provided a significant 
improvement of the fit, compared to any previously considered model 
($\chi^2=1318.3$ for 1360 dof; see Fig. \ref{xmmNuSTARsoft2} and 
Tab. \ref{xmmNuSoft}). 
The introduction of the ionised disc reflection component (green dotted 
line in Fig. \ref{xmmNuSTARsoft2}) allowed us to reproduce the broad 
excess in the Fe~K band well. Additionally, we note that the spectrum can be 
fitted with a cooler disc black body ($kT_{DBB}=1.55\pm0.11$~keV), 
with a slightly larger inner disc radius of $r_{DBB}=7.5\pm1.1$ km. 
Moreover, because of the Compton reflection hump, the reflection component 
has a harder spectrum in the 10-40 keV band compared to the illuminating source. 
This allowed us to reproduce part of the high energy emission in excess 
above the extrapolation of the disc black body component (see 
Fig. \ref{xmmNuSTARsoft2}). However, by fixing the temperature of 
the illuminating black body to the temperature of the disc black body, 
the ionised reflection could not reproduce all of the high energy emission, 
therefore it was still requiring the presence of a hot ($kT_{BB}=3.1\pm0.2$~keV) 
black body emission from small patches on the NS surface 
($r_{BB}\sim0.2-0.3$~km). 

We noted that the hot black body component could also contribute 
to the disc irradiation. If so, the effective temperature of the thermal 
emission irradiating the disc and producing the reflection spectrum 
could be higher than $kT_{DBB}$ (as assumed before). 
We therefore re-fitted the spectra with the same model, but assuming 
that the reflection is produced by a black body irradiation with temperature 
$kT_R=k(2\times T_{DBB}+T_{BB})/3$, intermediate between the temperature 
of the warm disc black body and hot black body one. 
The model, with these assumptions provided a significant worsening of 
the fit ($\Delta\chi^2=-30.2$ for the same dof). 
The central panel of Fig. \ref{xmmNuSTARsoft2} shows that, as expected, 
the reflection spectrum (green dotted line) is indeed harder than before. 
In particular, the spectral shape of the reflection component is similar 
to the one of the hot black body component (blue dash-dotted 
line). Such a hard reflection spectrum can therefore reproduce the majority 
of the hard X-ray emission, therefore it appears to be several times brighter 
than the hot black body emission. This seems physically unlikely. 
For this reason, we disregard this model and hereinafter we will assume 
$kT_R=kT_{DBB}$ (or $kT_R=kT_{BB}$). 

\subsubsection{Thermal plus Comptonisation and relativistic reflection}
\label{SoftThnthcompBBrefl}

\begin{figure}
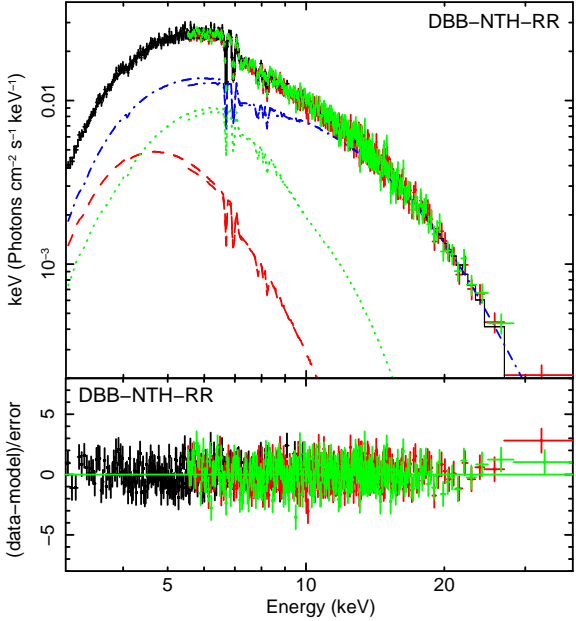

\includegraphics[height=0.45\textwidth,angle=-90]{DBB-NTH-RReuf.ps}
\vspace{-0.1cm}

\includegraphics[height=0.45\textwidth,angle=-90]{DBB-NTH-RRdel.ps}
\caption{\xmm+\nustar\ spectra of the longest simultaneous 
soft state observation (same as Fig. \ref{xmmNuSTARsoft}), 
once fitted with the model {\sc dbb-nth-rr}. The disk black body, 
Comptonisation ({\sc nthcomp}) and ionised reflection ({\sc bbrefl}) are 
shown with red dashed, blue dot-dashed and green dotted 
lines, respectively. }
\label{xmmNs}
\end{figure}
We then employed the same model, explored in section \ref{2thbbdl}, 
but we exchanged the {\sc diskline} component with a relativistic ionised 
reflection one, i.e., {\sc softabs*(diskbb+nthComp+kdblur(bbrefl))} and
{\sc softabs*(bbody+nthComp+kdblur(bbrefl))} and we called these 
{\sc dbb-nth-rr} and {\sc bb-nth-rr}, respectively (Tab. \ref{xmmNuSoft} 
and Fig. \ref{xmmNs}). 
We note that in all cases, the models involving the relativistic ionised 
reflection component provided a significantly better description of the data, 
compared to the ones employing a disk-line (Tab. \ref{xmmNuSoft}). 
Moreover, model {\sc dbb-nth-rr} appears preferred over the {\sc bb-nth-rr}
one ($\Delta\chi^2=21.3$ for the same dof). 

The model {\sc dbb-nth-rr} provided a good description of the data with 
reasonable best fit parameters. In particular, we note that the temperature 
of the disk black body component is significantly lower 
($kT_{DBB}=1.1\pm0.1$~keV) and the inner disc radius 
($r_{DBB}\sim9\pm3$~km) larger than with the other models considered. 
Indeed, the hard X-ray flux is now reproduced by the combination of 
the harder reflection and Comptonization components (see Fig. \ref{xmmNs}), 
requiring no strong contribution from the soft disk black body component. 
We observed that the un-absorbed flux of the disk black body emission 
in the 1-500~keV band is $F_{1-500}\sim1.8\times10^{-10}$ 
erg cm$^{-2}$ s$^{-1}$ about half the flux of the Comptonisation 
component $F_{1-500}\sim3.7\times10^{-10}$ erg cm$^{-2}$ s$^{-1}$, 
while the flux of the ionised reflection component is 
rather high being $F_{1-500}\sim1.5\times10^{-10}$ erg cm$^{-2}$ s$^{-1}$,
$\sim27$~\% of the total illuminating flux. 

\subsubsection{Three components model}

We finally fit the spectrum with the three components model 
({\sc dbb-bb-nthcomp-rr}), composed by the double thermal plus Comptonisation 
and relativistic ionised reflection ({\sc softabs*(diskbb+bbody+nthComp+kdblur(bbrefl))}). 
This model provides a significant improvement of the fit compared to previously 
tested ones ($\Delta\chi^2=18.5$ for the addition of 2 parameters, F-test 
probability $\sim8\times10^{-5}$). 

\begin{figure}
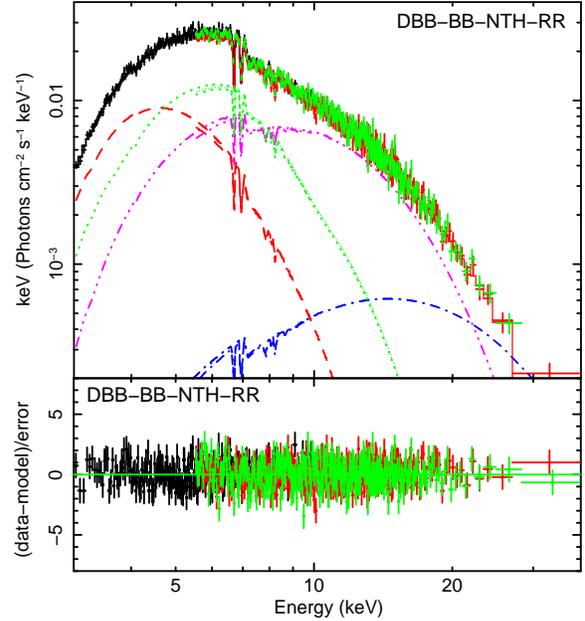

\includegraphics[height=0.45\textwidth,angle=-90]{DBB-BB-NTH-RReuf.ps}
\vspace{-0.1cm}

\includegraphics[height=0.45\textwidth,angle=-90]{DBB-BB-NTH-RRdel.ps}
\caption{\xmm+\nustar\ spectra of the longest simultaneous 
soft state observation (same as Fig. \ref{xmmNuSTARsoft}), 
once fitted with the three component model {\sc dbb-bb-nth-rr}. 
The disk black body, black body, Comptonisation ({\sc nthcomp}) 
and relativistic ionised reflection ({\sc bbrefl}) are shown with red dashed, 
magenta dot-dot-dashed, blue dot-dashed and green dotted lines, 
respectively. }
\label{xmmNs3comp}
\end{figure}

With this model the soft X-ray emission is dominated by a warm 
$kT_{DBB}=1.1\pm0.1$~keV disc black body component with a reasonable 
inner disc radius of $r_{DBB}=14\pm2$~km, corresponding to $\sim7$~r$_g$, 
therefore larger than the NS radius. 
A hotter black body component ($kT_{BB}=2.3\pm0.2$~keV) emitted from 
a small area with a radius $r_{BB}=0.6\pm0.2$~km that dominates 
the $\sim8-20$~keV band, while a faint and hard ($\Gamma\sim1-2.2$) 
Comptonisation component (with high energy cut off  at $kT_e\sim5$~keV) 
becomes important above $\sim20$~keV (see Fig. \ref{xmmNs3comp}). 

We observed that the un-absorbed flux of the disk black body emission 
in the 1-500~keV band is $F_{1-500}\sim3.9\times10^{-10}$ erg cm$^{-2}$ s$^{-1}$. 
A smaller, however comparable flux is observed to be produced by black body radiation 
$F_{1-500}\sim2.0\times10^{-10}$ erg cm$^{-2}$ s$^{-1}$, while the Comptonisation 
component carries only a small fraction of the energy 
$F_{1-500}\sim0.9\times10^{-10}$ erg cm$^{-2}$ s$^{-1}$. 
We note that the flux of the ionised reflection component is high 
$F_{1-500}\sim2.3\times10^{-10}$ erg cm$^{-2}$ s$^{-1}$, 
corresponding to $\sim30$~\% of the total. 

\subsection{\nustar+\xmm\ soft state: on 2015-04-01 and 2015-04-02} 

We then applied the same models best fitting the longest simultaneous 
\xmm+\nustar\ observation ({\sc dbb-bb-dl}, {\sc dbb-nth-rr}, {\sc bb-nth-rr} 
and {\sc dbb-bb-nth-rr}) to the other two soft state simultaneous \xmm\ 
plus \nustar\ observations. 
The statistics of these spectra is slightly lower. For this reason, 
the parameters (e.g., photon index and electron energy) of the weak Comptonised 
component can not be well constrained. Therefore we fitted these data assuming 
the best fit values observed during the 2015-02-25 observation 
($\Gamma=1.1$ and $kT_e=4.6$~keV). 

The fit of these additional two \xmm+\nustar\ spectra produced results 
similar to what was observed during the longer observation on 2015-02-26. 
Indeed, the spectra were best reproduced by the three component model
({\sc dbb-bb-nth-rr}). 
We also observed that the best fit parameters were consistent with the ones 
obtained during the 2015-02-25 observation (Tab. \ref{xmmNuSoftAltre}). 

The best fit un-absorbed flux of the disk black body, black body and Comptonised 
emission during the observation accumulated on 2015-04-02 were: 
$F_{1-200}\sim4.4\times10^{-10}$, $\sim1.9\times10^{-10}$ 
and $\sim0.1\times10^{-10}$ erg cm$^{-2}$ s$^{-1}$, respectively. 
The flux of the ionised reflection component was $\sim25$~\% of the total. 
Instead, on  2015-04-02 they were: 
$F_{1-200}\sim3.2\times10^{-10}$, $\sim2.7\times10^{-10}$ 
and $\sim0.7\times10^{-10}$ erg cm$^{-2}$ s$^{-1}$, respectively. 
The flux of the ionised reflection component was $\sim33$~\% of the total. 

\begin{table*}
\small
\begin{tabular}{ l l c c c c c c c c c c c c c c c}
\hline
\hline
\multicolumn{9}{c}{\bf Simultaneous \xmm+\nustar\ soft state} \\
& \multicolumn{4}{c}{\bf (2015-04-02)} & \multicolumn{4}{c}{\bf (2015-04-01)} \\
\hline
Model     &{\sc dbb-bb-rr}&{\sc dbb-nth-rr}&{\sc bb-nth-rr}&{\sc dbb-bb-nth-rr}&{\sc dbb-bb-rr}&{\sc dbb-nth-rr}&{\sc bb-nth-rr}&{\sc dbb-bb-nth-rr}& \\
$N_{H}$         &$31.5\pm0.6$ &$31.3\pm0.9$    &$27.0\pm0.6$     &$31.5\pm0.9$     &$32.8\pm1.0$ &$31.1\pm1.3$&$28.4\pm1.2$  &$31.5\pm1.2$  \\
$log(\xi_{IA})$ &$4.1\pm0.2$  &$4.1\pm0.2$     &$4.1\pm0.2$      &$4.1\pm0.2$      &$3.9\pm0.3$  &$3.9\pm0.3$ &$4.3\pm0.4$   &$3.9\pm0.4$   \\
$log(N_{H_{IA}})$&$23.6\pm0.3$ &$23.7\pm0.3$    &$23.6\pm0.3$     &$23.6\pm0.3$     &$23.3\pm0.4$ &$23.3\pm0.3$&$23.6\pm0.5$  &$23.2\pm0.5$  \\
$kT_{DBB}$      &$1.3\pm0.1$  &$1.2\pm0.1$     &                 &$1.2\pm0.2$      &$1.4\pm0.3$  &$1.05\pm0.4$ &             &$1.02\pm0.15$  \\
$N_{DBB}$       &$8.8\pm2.4$  &$8.8\pm2.1$     &                 &$12\pm3$         &$9\pm4$      &$7\pm5$      &             &$16\pm3$       \\
$kT_{BB}$       &$2.8\pm0.2$  &                &$1.08\pm0.07$    &$2.6\pm0.2$      &$2.7\pm0.3$  &             &$1.0\pm0.3$  &$2.2\pm0.1$   \\
$N_{BB}$        &$0.2\pm0.1$  &                &$11\pm5$         &$0.4\pm0.2$      &$0.3\pm0.2$  &             &$6\pm5$      &$0.9\pm0.3$  \\
$\Gamma$       &             &$1.4\pm0.4$     &$1.4\pm0.5$      &$1.1\dag$        &             &$1.7\pm0.1$  &$1.7\pm0.2$  &$1.1\dag$   \\
$kT_{e}$        &             &$2.9\pm0.2$     &$2.8\pm0.2$      &$4.5\dag$        &             &$2.9\pm0.1$  &$2.9\pm0.2$  &$3.4\dag$  \\
$N_{nth}$       &             &$0.7^{+1.3}_{-0.6}$&$0.3^{+0.4}_{-0.2}$ &$0.003\pm0.002$  &             &$3.2\pm1.3$  &$0.8\pm0.3$  &$0.01\pm0.01$  \\
$N_{ref}$       &$76\pm14$    &$76\pm13$       &$73\pm16$        &$85\pm15$        &$63\pm30$   &$83\pm22$    &$75\pm15$     &$100\pm28$   \\
$c_{NuA}$       &$1.09\pm0.02$&$1.09\pm0.01$   &$1.09\pm0.02$    &$1.09\pm0.02$    &$1.11\pm0.02$&$1.11\pm0.02$&$1.11\pm0.02$&$1.11\pm0.02$\\
$c_{NuB}$       &$1.12\pm0.02$&$1.12\pm0.01$   &$1.12\pm0.02$    &$1.12\pm0.02$    &$1.20\pm0.02$&$1.20\pm0.02$&$1.20\pm0.02$&$1.20\pm0.02$\\
$\chi^2/dof$   &1284.8/1271  &1283.6/1270     &1304.4/1270      &1280.6/1270      &1257.5/1306  &1245.6/1305  &1249.8/1305  &1241.6/1305  \\
\hline
\hline
\end{tabular} 
\caption{Best fit parameters for the two soft state simultaneous \xmm+\nustar\ 
observations accumulated on 2015-04-02 and 2015-04-1. 
See the caption of Tab. \ref{xmmNuSoft} for a description of the parameters. }
\label{xmmNuSoftAltre}
\end{table*} 

\subsection{\nustar+\xmm\ hard state} 

We then fitted the two hard state contemporaneous \xmm\ plus \nustar\ spectra 
(Tab. \ref{xmmNuHard}). Each of the two long \nustar\ exposures (which 
started on 2014-08-30 and on 2014-09-27) was partially covered by 
two shorter \xmm\ observations (Tab. \ref{data}). For this reason, 
we fitted the four spectra (2 \xmm\ plus \nustar\ FPMA and FPMB) together. 
We observed no major spectral variation between the \xmm\ and \nustar\ 
observations, therefore we employed the same 
model, with the same parameters, to fit the four spectra. We only allowed 
a normalisation constant to be different between the various spectra 
(in order to reproduce small normalisation differences, 
associated with minor source flux variations; see Tab. \ref{xmmNuHard}). 
\begin{table*}
\small
\begin{tabular}{ l c c c c c c c c c c c c c c c}
\hline
\hline
\multicolumn{6}{c}{\bf Simultaneous \xmm+\nustar\ hard state (2014-09-28)}   \\
\hline
Model        & PL           & {\sc simpdbb}  &{\sc simpbb} & \multicolumn{2}{c}{\sc dbb-nth}   &  \\
$N_{H}$       &$28.5\pm1.2$  &$28.8\pm1.3$    &$-$          &$27.8\pm1.2$  &$30.5\pm2.0$  &  \\
$kT_{DBB}$    &              &$0.2\star$      &             &$0.2$         &$0.7$           \\
$N_{DBB}$     &              &$720\pm30$      &             &$<170000$     &$6\pm3$         \\
$kT_{BB}$     &              &                &$0.4\star$   &              &                \\
$N_{BB}$      &              &                &$97\pm2$     &              &                \\
$\Gamma$     &$1.83\pm0.05$ &$1.84\pm0.03$   &$-$          &$1.84\pm0.02$ &$1.84\pm0.02$   \\
$N$          &$1.34\pm0.08$ &                &             &              &                \\
$F_{sc}$      &              &$1.0\star$      &$1.0\star$   &              &                \\
$N_{Nu}$      &$1.3\pm0.1$   &                &             &              &                 \\
$kT_{e}$      &              &                &             &$>120$        &$>100$           \\
$N_{nth}$     &              &                &             &$1.29\pm0.07$ &$0.76\pm0.03$    \\
$c_{xmm2}$    &$0.98\pm0.02$ &$0.97\pm0.01$   &$-$          &$0.97\pm0.01$ &$0.97\pm0.02$    \\
$c_{NuA}$     &              &$0.95\pm0.03$   &$-$          &$0.93\pm0.03$ &$0.91\pm0.03$    \\
$c_{NuB}$     &$1.10\pm0.02$ &$1.04\pm0.03$   &$-$          &$1.02\pm0.03$ &$1.00\pm0.03$    \\
$\chi^2/dof$ &2360.2/2218   &2361.7/2220     &2360.7/2220  &2380.7/2218   &2354.9/2218      \\
\hline
\hline
\multicolumn{6}{c}{\bf Simultaneous \xmm+\nustar\ hard state (2014-08-31)}  \\
\hline
Model        & PL           & {\sc simpdbb}  &{\sc simpbb}  & \multicolumn{2}{c}{\sc dbb-nth} &    \\
$N_{H}$       &$28.3\pm1.1$  &$28.1\pm1.2$      &$-$             &$28.0\pm0.6$  &$31.4\pm2.5$  &  \\
$kT_{DBB}$    &              &$0.2\star$        &                &$0.2$         &$0.7$           \\
$N_{DBB}$     &              &$710\pm35$        &                &$<160000$     &$9\pm4$       \\
$kT_{BB}$     &              &                  &$0.4\star$      &              &                \\
$N_{BB}$      &              &                  &$90\pm2$        &              &                \\
$\Gamma$     &$1.88\pm0.03$ &$1.88\pm0.03$     &$-$             &$1.88\pm0.02$ &$1.87\pm0.02$   \\
$N$          &$1.36\pm0.08$ &                  &                &              &                \\
$F_{sc}$      &              &$1.0\star$        &$1.0\star$      &              &                \\
$N_{Nu}$      &$1.41\pm0.11$ &                  &                &              &                 \\
$kT_{e}$      &              &                  &                &$>150$        &$>70$           \\
$N_{nth}$     &              &                  &                &$1.34\pm0.07$ &$0.72\pm0.03$    \\
$c_{xmm2}$    &$1.09\pm0.02$ &$1.09\pm0.02$     &$-$             &$1.09\pm0.02$ &$1.09\pm0.02$    \\
$c_{NuA}$     &              &$1.02\pm0.03$     &$-$             &$1.02\pm0.03$ &$1.02\pm0.03$    \\
$c_{NuB}$     &$1.07\pm0.02$ &$1.09\pm0.03$     &$-$             &$1.09\pm0.04$ &$1.08\pm0.03$    \\
$\chi^2/dof$ &2174.7/2130   &2187.1/2130       &2184.6/2130     &2191.2/2128   &2164.9/2128      \\
\hline
\hline
\end{tabular} 
\caption{Best fit parameters for the two hard state simultaneous \xmm+\nustar\ 
observations accumulated on 2014-09-28 and 2014-08-31. 
$c_{xmm2}$ shows the normalisation constant of the second 
\xmm\ spectrum. See as caption of Tab. \ref{xmmNuSoft} for a description 
of the other parameters. 
$\star$ Indicated the parameter is fixed to the given value. For model {\sc simpdbb}, 
we report the variation of the best fit parameters to the assumed parameters. 
{\it For the observation accumulated on 2014-09-28.} 
For example, assuming $F_{sc}=1$ and $kT_{DBB}=0.4,0.7$ and $1.0$~keV, 
we obtained $N_{DBB}=46\pm2, 5.1\pm0.1$ and $1.27\pm0.03$, with 
$\chi^2/dof=2361.5/2220, 2363.2/2220$ and $2385.9/2220$, respectively 
(all other parameters of the model do not vary significantly), 
while assuming $kT_{DBB}=0.4$~keV and $F_{sc}=0.6$ and $0.3$, we obtained 
$N_{DBB}=78\pm3$ and $160\pm5$, with $\chi^2/dof=2362.5/2220$ and 
$2366.2/2220$. We then performed the same exercise for model {\sc simpbb}. 
For $kT_{BB}=0.7$ and $F_{sc}=1.0$, we obtain $N_{BB}=10.9\pm0.2$ 
and $\chi^2/dof=2383.4/2220$, while for $kT_{BB}=0.4$ and $F_{sc}=0.3$, 
we obtain $N_{BB}=362\pm8$ and $\chi^2/dof=2381.0/2220$. 
{\it For the observation accumulated on 2014-08-31.} 
For example, assuming $F_{sc}=1$ and $kT_{DBB}=0.4,0.7$ and $1.0$~keV, 
we obtained $N_{DBB}=44\pm2, 4.8\pm0.1$ and $1.19\pm0.03$, with 
$\chi^2/dof=2186.6/2130, 2188.1/2130$ and $2214.0/2130$, respectively 
(all other parameters of the model do not vary significantly), 
while assuming $kT_{DBB}=0.4$~keV and $F_{sc}=0.6$ and $0.3$, we obtained 
$N_{DBB}=74\pm3$ and $150\pm6$, with $\chi^2/dof=2188.8/2130$ and 
$2194.9/2130$. We then performed the same exercise for model {\sc simpbb}. 
For $kT_{BB}=0.7$ and $F_{sc}=1.0$, we obtain $N_{BB}=10.0\pm0.2$ 
and $\chi^2/dof=2209.7/2130$, while for $kT_{BB}=0.4$ and $F_{sc}=0.3$, 
we obtain $N_{BB}=340\pm9$ and $\chi^2/dof=2210.6/2130$. }
\label{xmmNuHard}
\end{table*} 

We started by fitting the \xmm\ plus \nustar\ spectra with the absorbed 
power-law model best fitting the \xmm\ spectra (\S \ref{xmmhard}). 
This simple model very well reproduces the X-ray emission of \axj, 
during the hard state. The observed power law photon index 
($\Gamma=1.86-1.88$) is, in all cases, within the range of values 
typically observed during the hard state. The inter-normalisation constants 
are within $\sim10$~\% of the expected value, indicating a minor 
flux variation of the source between the different spectra. 

As typically observed in accreting BH and NS, the disc black 
body emission is far less prominent during the hard state than 
in the soft (Done et al. 2007; Dunn et al. 2010). 
In fact, during hard states, the thermal emission is commonly 
weaker and far colder than during the soft state, with observed 
temperatures in the range: $kT\sim0.2-0.7$~keV (Dunn et al. 2010). 
Being \axj\ highly absorbed, we can not 
detect any X-ray photon below $\sim3$~keV, preventing us from placing 
strong constraints on the temperature of such thermal component
during the hard state. 
On the other hand, to reconstruct fiducial source SED it is essential 
to constrain the position and intensity of the disc emission. 
Therefore, to restrain this, we fitted the spectra with an array 
of models containing a soft thermal component, in addition 
to the Comptonisation one, producing the power law emission. 
In particular, we employed the models {\sc simpdbb} 
and {\sc simpbb} that are reminiscent of the same models 
that we used for the soft state, once removing the disk-line and 
ionised absorption components (resulting in the models: 
{\sc hardabs*(simpl*diskbb)}) and {\sc hardabs*(simpl*bbody)}, 
respectively). For each of these models we explored an array 
of possible parameters, such as disk black body temperatures 
and/or Comptonisation fractions\footnote{We define as 
Comptonisation fraction the fraction of seed photons that scatter 
producing inverse Compton radiation (such as defined in the {\sc simpl} 
model; Steiner et al. 2009). } (Tab. \ref{xmmNuHard}). 

We observed that any disk black body component with 
temperature in the range $kT_{DBB}\sim0.2-0.7$~keV and 
Comptonisation fractions within $f_{sc}\sim0.3-1.0$ are 
consistent with the data, while hotter $kT_{DBB}\sim1.0$~keV 
(or $kT_{BB}\sim0.7$~keV) thermal emission produces 
a significantly worse fit (Tab. \ref{xmmNuHard}). 

To constrain the presence of a high energy cut off, 
we also employed the model {\sc dbb-nth} (implemented 
in \xspec\ as {\sc hardabs*(diskbb+nthcomp)} ). For both series of data-sets, 
we observed either better or comparable results 
whenever the disk black body temperature is assumed to be 
in the range $kT_{DBB}\sim0.2-0.7$~keV (Tab. \ref{xmmNuHard}). 
For disk temperatures significantly higher than $kT_{DBB}=0.7$~keV, 
the inner disc radius results to be smaller than the NS radius, 
therefore we can rule this out as unphysical. On the other 
hand, smaller disk temperatures are possible and they imply 
only upper limits to the inner disc radius (e.g., for 
$kT_{DBB}=0.2$~keV, $r_{DBB}<10^3$~km; Tab. 
\ref{xmmNuHard}). 
No high energy cut off is detected within the observed X-ray band. 
Indeed, only lower limits to the temperature of the Comptonising 
electrons is observed ($kT_e>70-100$~keV; Tab. \ref{xmmNuHard}). 
This appears very different to what is observed during the soft state, 
when the cut off is at $kT_e\sim3-5$~keV (\S 5.1). 
We conclude by observing that this simplified model can 
reproduce the observed X-ray spectra during the hard state 
of \axj.

\section{Fit of the remaining \nustar\ observations}

\begin{table*}
\small
\begin{tabular}{ l l c c c c c c c c c c c c c c c}
\hline
\hline
\multicolumn{10}{c}{\bf \nustar} \\
& \multicolumn{5}{c}{\bf Soft state} &  \multicolumn{2}{c}{\it Hard state} \\
               & 2015-03-31       & 2013-08-23      & 2013-08-13         & 2013-08-09        & 2013-08-08     & {\it 2014-06-18}& {\it 2013-07-31} & &\\
\hline
Model       &{\sc dbb-bb-nth-rr}&{\sc dbb-bb-nth-rr}&{\sc dbb-bb-nth-rr}&{\sc dbb-bb-nth-rr}&{\sc dbb-bb-nth-rr}&{\sc dbb-nth}  & {\sc dbb-nth} & \\
$kT_{DBB}$      &$1.5^{+0.1}_{-0.4}$ &$1.2\pm0.2$      &$1.5^{+0.1}_{-0.3}$&$1.5^{+0.2}_{-0.4}$&$1.6^{+0.2}_{-0.4}$&             &             \\
$N_{DBB}$       &$7^{+4}_{-1}$      &$15\pm10$        &$6^{+6}_{-1}$     &$6^{+15}_{-2}$    &$3^{+17}_{-1}$    &              \\
$kT_{BB}$       &$3.0\pm0.4$      &$2.3\pm0.2$      &$2.7\pm0.3$     &$2.5\pm0.4$     &$2.5\pm0.3$    &             &              \\
$r_{BB}$       &$0.11^{+0.12}_{-0.02}$&$0.6\pm0.2$      &$0.2^{+0.3}_{-0.1}$&$0.3^{+0.4}_{-0.2}$&$0.3\pm0.2$    &             &              \\
$\Gamma$       &$1.1$            &$1.1$             &$1.1$          &$1.1$           &$1.1$           &$1.94\pm0.04$&$1.93\pm0.05$ \\
$kT_{e}$        &$4.5$            &$4.5$             &$4.5$          &$4.5$           &$4.5$           &$>90$        &$>90$        \\
$N_{nth}$       &$<0.01$           &$0.009\pm0.003$  &$<0.01$        &$0.010\pm0.005$ &$<0.01$         &$2.6\pm0.2$  &$3.3\pm0.02$  \\
$N_{ref}$       &$33^{+15}_{-9}$     &$42^{+33}_{-21}$    &$19^{+20}_{-10}$ &$17^{+10}_{-4}$    &$15\pm15$       &             &               \\
$c_{NuB}$       &$1.03\pm0.01$     &$1.03\pm0.01$    &$1.02\pm0.01$  &$1.04\pm0.02$   &$1.03\pm0.02$    &$1.09\pm0.02$&$1.02\pm0.02$  \\
$\chi^2/dof$   &542.1/634         &735.3/700        &560.4/519      &496.8/514       &578.8/519        &909.1/837   &839.4/772      \\
\hline
\hline
\end{tabular} 
\caption{Best fit parameters for the soft and hard state \nustar\ 
only observations. See as caption of Tab. \ref{xmmNuSoft} for 
a description of the various parameters. 
The ionisation parameter and column density of the ionised 
absorption have been fixed to $log(\xi_{IA})=4.1$ and 
$log(N_{H_{IA}})=23.5$, respectively. The column density of the neutral 
absorber is fixed to $N_{H}=32\times10^{22}$~cm$^{-2}$. The photon index 
and electron temperature of the Comptonised component are fixed to 
$\Gamma=1.1$ and $kT_e=4.5$~keV.}
\label{NuSTARonly}
\end{table*} 
We then fitted all soft state \nustar\ only observations with the best fit 
three component model {\sc dbb-bb-nth-rr}. Because of the lower
statistics (compared to the simultaneous \xmm+\nustar\ fits), 
we fixed the equivalent hydrogen column density, the column density 
and ionisation parameter of the ionised absorber as well as the photon 
index and electron temperature of the Comptonisation component to 
the best fit values observed during the 2015-02-25 observation
($N_{H}=32\times10^{22}$~cm$^{-2}$, $log(\xi_{IA})=4.1$,
$log(N_{H_{IA}})=23.5$, $\Gamma=1.1$ and $kT_e=4.5$~keV).
We observed that this model always provides a good description 
of the data, with best fit parameters similar to the one derived from 
the simultaneous \xmm+\nustar\ observations (Tab. \ref{NuSTARonly}). 

We also modelled the hard state \nustar\ only observation with 
the model {\sc dbb-nth}, also in this case obtaining acceptable fits 
and reasonable best fit parameters (see Tab. \ref{NuSTARonly}).

\section{Discussion}

We analysed the persistent emission from \axj. 
Thanks to simultaneous \nustar\ + \xmm\ observations 
we could detail the emission mechanisms and obtain 
accurate X-ray SEDs for both the soft and hard state. 
We note that, previous spectral fitting works on \axj\ 
were restricted to the limited energy band provided by \xmm\ 
(Ponti et al. 2015), therefore impeding detailed fits of the X-ray 
continuum with models more complex than a simple power law 
or (disk)black body emission for the hard and soft states, 
respectively. Thanks to the addition of the NuSTAR data 
we demonstrated the richness of information of the X-ray 
emission from \axj. 

Strong spectral and flux variability is observed between  
the two states, on the other hand only moderate variations 
are observed between different observations within 
the same state. 

All observations show clear evidence for a high column 
density of neutral absorption ($N_H\sim3.0\times10^{23}$~cm$^{-2}$). 
While it is well known that in dipping sources (during dips) the 
neutral absorption is highly variable (Frank et al. 1987; Diaz-Trigo 
et al. 2006; Ponti et al. 2016), we observed that all soft and hard 
state spectra are roughly consistent with the same column density 
of absorbing material, once they are fitted with the same model 
(e.g., model {\sc dbb-bb-nth-rr} and {\sc simpbb}). 
This is consistent with the idea that a significant fraction of the neutral 
absorption (during persistent emission) is produced by the interstellar 
medium (see Jin et al. 2017a,b; Ponti et al. 2017). 

\subsection{Soft state observations}
All soft state observations of \axj\ show clear signs of ionised 
absorption lines in the \xmm\ and \nustar\ spectra, signatures 
of a highly ionised ($log(\xi_{IA})\sim4.1$) and high column density 
($log(N_{H_{IA}})\sim23.5-24$) plasma. We note that these values 
are within the range of typical ionisation states and column densities 
observed in NS and BH LMXB (King et al. 2013; Ponti et al. 2016; 
Diaz-Trigo et al. 2016). Because of the limited energy resolution, 
we can only place weak constraints on the bulk outflow velocity 
of this plasma ($v_{out}\leq2000$~km~s$^{-1}$). 
The ionised plasma is consistent with being constant within all 
soft state observations. 

During the soft state, once the neutral and ionised absorptions 
are reproduced, broad positive residuals appear in 
the $\sim5-7$~keV band. We explored whether such excess could 
be the manifestation of the emission component of a P-Cygni profile. 
Nevertheless, the broadness of this feature is inconsistent with 
the outflow velocities typically observed in winds in NS and 
constrained in \axj. 
Alternatively, we explored whether such feature might be reproduced 
by a broad Fe~K$\alpha$ emission line, reflected off the accretion disc. 
The introduction of a disk-line profile provides an acceptable fit, 
with a line equivalent width $EW\sim120-200$~eV, consistent with 
disc reflection (Matt et al. 1993). The result for the best fit disc 
inclination is high, roughly consistent with the eclipsing behaviour of 
the source, and the disc emissivity 
index is scaling with radius as $r^{-2.4\pm0.1}$, a value 
consistent with illumination of a flared disc by a central source. 
We note that for all combinations of continuum emission models 
the substitution of the disk-line component with a self consistent 
ionised reflection spectrum provides a significant improvement 
of the fit. Although we cannot rule out alternative hypotheses 
(e.g., more complex absorption, etc.), this strengthens the suggestion 
that the positive excess in the $5-7$~keV band has an origin 
as reprocessing from the accretion disc. 

The soft X-ray spectrum (within the \xmm\ band) of \axj\ is dominated 
by a prominent thermal component. 
The three emission component ({\sc dbb-bb-nth-rr}) model 
provides a superior fit and an 
excellent description of all soft state spectra. 
In this model, the continuum is produced by a prominent 
disk black body component with a temperature of 
$kT_{DBB}=1.0-1.2$~keV, dominating the emission 
below $\sim5$~keV. Such a range of temperatures is 
theoretically expected and typically observed in BH and 
NS in the soft state. Assuming that \axj\ is located at the 
GC and that its disc is highly inclined 
($i\sim80^\circ$, indeed it shows eclipses) and some 
correction factors (see Kubota et al. 1998), 
the best fit inner disc radius results to be $r_{DBB}\sim12-16$~km 
($\sim7$~r$_g$). As expected, this value is comparable (however 
larger) than the typical NS radius. In the best fit model ({\sc 
dbb-bb-nth-rr}), the second emission component, required to reproduce 
the hard X-ray emission measured by \nustar, is associated 
with black body emission, possibly connected to the boundary layer 
at the NS surface. This component dominates the source emission 
in the $\sim8-20$~keV band. Its temperature ranges within 
$kT_{BB}\sim2.2-3.0$~keV with an associated very small emitting 
radius of $r_{BB}\sim0.5-0.8$~km. In theory, this radiation might 
be associated with emission from small and hot patches 
or an equatorial bundle on the NS surface (e.g., where 
the accretion column impacts the NS surface). 
The emission above $\sim20$~keV can be fit by the addition 
of a faint Comptonisation component. In this three component 
model, the Comptonisation emission is very weak, preventing a 
detailed characterisation of its parameters. The best fit 
asymptotic photon index and electron temperature are 
$\Gamma=1.1^{+1.1}_{-0.2}$ and $kT_e=4.6^{+10.3}_{-2.4}$ keV, 
respectively. These values carry large uncertainties, however 
they indicate a significant optical thickness of this medium with 
a scattering optical depths ($\tau$) of $\tau\geq3$ (assuming the relation 
$\Gamma=[9/4 + \frac{1}{(kT_e/m_ec^2)\tau(1+\tau/3)} ]^{1/2} -1/2$;
Sunyaev \& Titarchuk 1980). 
Finally, the addition of a relativistic ionised reflection model also 
improves the fit. We observed that in all cases, the combination of 
disk black body plus black body emission carry most ($\sim70-75$~\%) 
of the flux within the 1-200~keV band and the ratio of black body 
to disk black body radiation is $\sim0.4-0.8$. 
On the other hand the Comptonisation component carries only 
$\sim10$~\% of the 1-200~keV emission, while a significant contribution 
($\sim20-30$~\%) is provided by the relativistic ionised reflection. 

We note, that in the three component model, the bulk of 
the $\sim8-20$~keV emission is produced by the black body radiation 
and the Comptonised component is very faint, and relegated to higher 
energies. 
Both the spectral parameters and the relative contributions of 
the emission components are consistent with those observed 
in a very similar system, 4U~1608-52, during the soft state and using 
the same modelling (Armas Padilla et al. 2017). This system is seen 
through an $N_\mathrm{H}$ 20--30 times lower than that of \axj, 
which enables to study the softest spectral region ($<$3 keV).  
We also note that the low comptonization fractions inferred from 
the three component model are systematically seen in BH soft 
states (e.g. Dunn et al. 2011, Mu\~noz-Darias et al. 2013). 
This behaviour seems reasonable given that both BH and NS systems 
reach similarly low fast variability levels during the soft state, 
which is likely to be produced in the comptonization component 
(e.g.  Mu\~noz-Darias et al. 2014; see also Lin et al. 2007, 2009).

Alternatively, the soft state spectra are reasonably well 
reproduced by a two component model (e.g. disk black body emission plus 
comptonisation). In this case, the comptonised component would be more prominent, 
producing most of the emission above $8$~keV. If so, a high energy 
cut off must occur within the X-ray band, implying that the 
comptonised emission is produced by a population of low temperature 
electrons ($kT_e\sim3-4$~keV). 

We stress, that this electron temperature 
is one to two orders of magnitude smaller than what is typically observed 
in BH binaries (both soft and hard states) and in NS during the hard state. 
In this scenario, the best fit photon index is $\Gamma=1.7-2.2$, therefore 
consistent with the values typically observed in accreting BH and NS. 
The observed photon index and electron temperature imply a very high 
optical depth of this plasma ($\tau\sim8-12$). This means that, even when 
the high energy radiation is fitted with a (non-thermal) comptonisation 
model, the comptonised radiation goes into the limit of becoming 
nearly-thermal black body radiation, right where the comptonisation 
radiation and black body blur. 

\subsection{Hard state observations}

During the hard state, \axj's X-ray emission is dominated by a power 
law with photon index $\Gamma\sim1.8-2.0$, showing no evidence for 
a high energy cut off ($kT_e>70-100$~keV). This implies a small 
optical depth of the comptonising plasma ($\tau<1.6$). 
Would the high energy cut off be located at $kT_e=300$ or 
$kT_e=800$~keV, then the comptonising plasma optical depth 
would be $\tau\sim0.4-0.15$, resulting into a very 
optically thin layer. The large lower limit to the energy of the 
cut off appears rather high compared to what is typically 
observed in accreting NS (Burke et al. 2017) and this might be 
related to inclination effects (Makishima et al. 2008; Zhang et al. 
2014). 

We also confirm previous results observing the disappearance 
of the ionised absorption plasma in the hard state. As already 
discussed in previous works, it is excluded that the absorption 
disappears because of over-ionisation, during the hard state 
(Ponti et al. 2015). 
To understand the origin of this variation, we constructed here detailed 
and accurate SED to use as input for investigating the photoionisation 
stability of the absorbing plasma. We will show in a companion paper 
that, whenever the ionised plasma observed during the soft state is illuminated 
by a hard state SED, it becomes unstable. Therefore, it has to change 
its physical conditions (Bianchi et al. 2017). 

Tight upper limits are observed also on the presence of broad 
emission lines ($EW<30$~eV), during the hard state. 
This could be the consequence of a major variation in the accretion 
flow, with the optically thick layer accretion disc that is ubiquitous 
during the soft state becoming optically thin (therefore producing 
no reflection component) inside a (rather large) truncation radius, 
during the hard state (Done et al. 2007; 
Plant et al. 2014; 2015; De Marco et al. 2015a,b; 2016). 

\section{Conclusions}

We presented 11 new \xmm\ observations as well as 15 new \nustar\ 
data-sets, that caught \axj\ in outburst, therefore building a large 
database (of almost 40 observations) of good resolution X-ray spectra 
of a high inclination NS X-ray binary accreting at intermediate rates (i.e. atoll regime). 

\begin{itemize}

\item{} We built accurate X-ray SEDs, representative of each state,
starting from the best fit models best reproducing the 
\xmm\ and \nustar\ spectra. We also reported radio (\gmrt) and 
optical (\grond) upper limits that are consistent with the known radio 
and optical to X-ray relations. 

\item{} All soft state observations are well described by a three 
component model. The best fit is provided by a disc black body component 
with $kT_{DBB}\sim1.1-1.2$~keV an inner disc radius 
$r_{DBB}\sim12-16$~km~$\sim7$~$r_g$, plus a hot 
$kT_{BB}\sim2.2-3.0$~keV black body component with 
a small emitting radius $r_{BB}\sim0.5-0.8$~km, possibly produced 
by the boundary layer at the NS surface, plus a faint comptonisation. 
Additionally, neutral plus ionised absorption and relativistic ionised 
reflection components are required by the data. 

\item{} All hard state observations are dominated by hard X-ray 
radiation, well reproduced by a rather flat power law emission 
($\Gamma\sim1.8-2.0$). No significant curvature is detected 
in the \xmm+\nustar\ band, indicating no requirement for a 
high energy cut off up to $\sim70-140$~keV. This implies a 
small optical depth of the comptonising plasma ($\tau<1.6$). 

\item{} We confirm, tripling the number of X-ray observations, 
the ubiquitous presence of Fe~K absorption lines during the soft state 
$log(\xi_{IA})\sim3.7-4.3$, $Log(N_{H_{IA}})\sim23.4-23.5$, 
$v_{out}<2000$~km~s$^{-1}$ and $v_{turb}\sim500-700$~km~s$^{-1}$. 
The plasma physical parameters remain roughly constant during all 
the soft state observations, while the ionised absorption features 
are significantly weakening during the hard state, 
as observed in archival data (Ponti et al. 2015). We will investigate
the dependence of the plasma properties on the source SED (therefore 
on its photo-ionisation stability) in a companion paper (Bianchi et al. 2017). 

\item{} During all soft state observations positive residuals 
remain in the $\sim6-8$~keV band. Such emission can be well 
reproduced by a reflection component with an Fe K$\alpha$ line  
with $EW\sim120-200$~eV. The best fit parameters of the broad 
emission line indicate a rather standard disc emissivity 
($r^{-2.4\pm0.1}$) from a highly inclined accretion disc 
($\alpha=70^{+7}_{-15}$$^{\circ}$). 

The disc line is not observed during the hard state, in line with 
the idea that the disc might be truncated, during the hard state 
(e.g. Plant et al. 2014; 2015; De Marco et al. 2015; 2016). 

\item{} Although the ionised absorption is highly variable between 
observations in the soft and in the hard state, a constant column 
density of neutral absorption can fit all \xmm\ spectra and all 
\nustar\ spectra of the persistent (out of dip) emission of \axj. 
This would be expected if the majority of the neutral absorption 
is due to material in the interstellar medium (Jin et al. 2017a,b). 

\end{itemize}

\section*{Acknowledgments}

The authors wish to thank Jan-Uwe Ness, Karl Foster, Ignacio de la Calle and 
the rest of the \xmm\ and \nustar\ scheduling teams for the support that made 
the coordinated observations possible. We would like to thank the referee 
for the helpful comments and careful reading of the paper.
The GC \xmm\ monitoring project is supported by the Bundesministerium 
f\"{u}r Wirtschaft und Technologie/Deutsches Zentrum f\"{u}r Luft- und Raumfahrt 
(BMWI/DLR, FKZ 50 OR 1408 and FKZ 50 OR 1408) and the Max Planck Society. 
SB acknowledges financial support from the Italian Space Agency under 
grant ASI-INAF I/037/12/0. TMD acknowledges support via a Ram\'on y Cajal 
Fellowship (RYC-2015-18148). BDM acknowledges support from 
the European UnionÕs Horizon 2020 research and innovation programme 
and the Polish National Science Centre grant Polonez 2016/21/P/ST9/04025. 
These results are based on observations obtained with XMM-Newton, 
an ESA science mission with instruments and contributions directly 
funded by ESA Member States and NASA
This work made use of data from the \nustar\ mission, a project led by the California 
Institute of Technology, managed by the Jet Propulsion Laboratory, and funded 
by the National Aeronautics and Space Administration. 
We thank the \nustar\ Operations, Software and Calibration teams for support 
with the execution and analysis of these observations. This research has made 
use of the \nustar\ Data Analysis Software (NuSTARDAS) jointly developed 
by the ASI Science Data Centre (ASDC, Italy) and the California Institute 
of Technology (USA).
We thank the staff of the GMRT who have made these observations
possible. The GMRT is run by the National Centre for Radio 
Astrophysics of the Tata Institute of Fundamental Research. 
Part of the funding for GROND (both hardware as well as personnel) was 
generously granted from the Leibniz-Prize to Prof. G. Hasinger 
(DFG grant HA 1850/28-1).

\appendix

\section{Further details on \nustar\ data reduction}
\label{NDR}

\subsection{Bright transients}
\label{Trans}

The immense improvement in the \nustar\ point spread function (PSF), 
compared to previous hard X-ray telescopes, allows us to accurately 
study the hard X-ray emission from relatively faint sources like \axj\ 
in extremely crowded fields such as the GC (Mori et al. 2015; Hong et al. 2015). 

\begin{figure}
\hspace{-0.3cm}
\includegraphics[height=0.445\textwidth,angle=0]{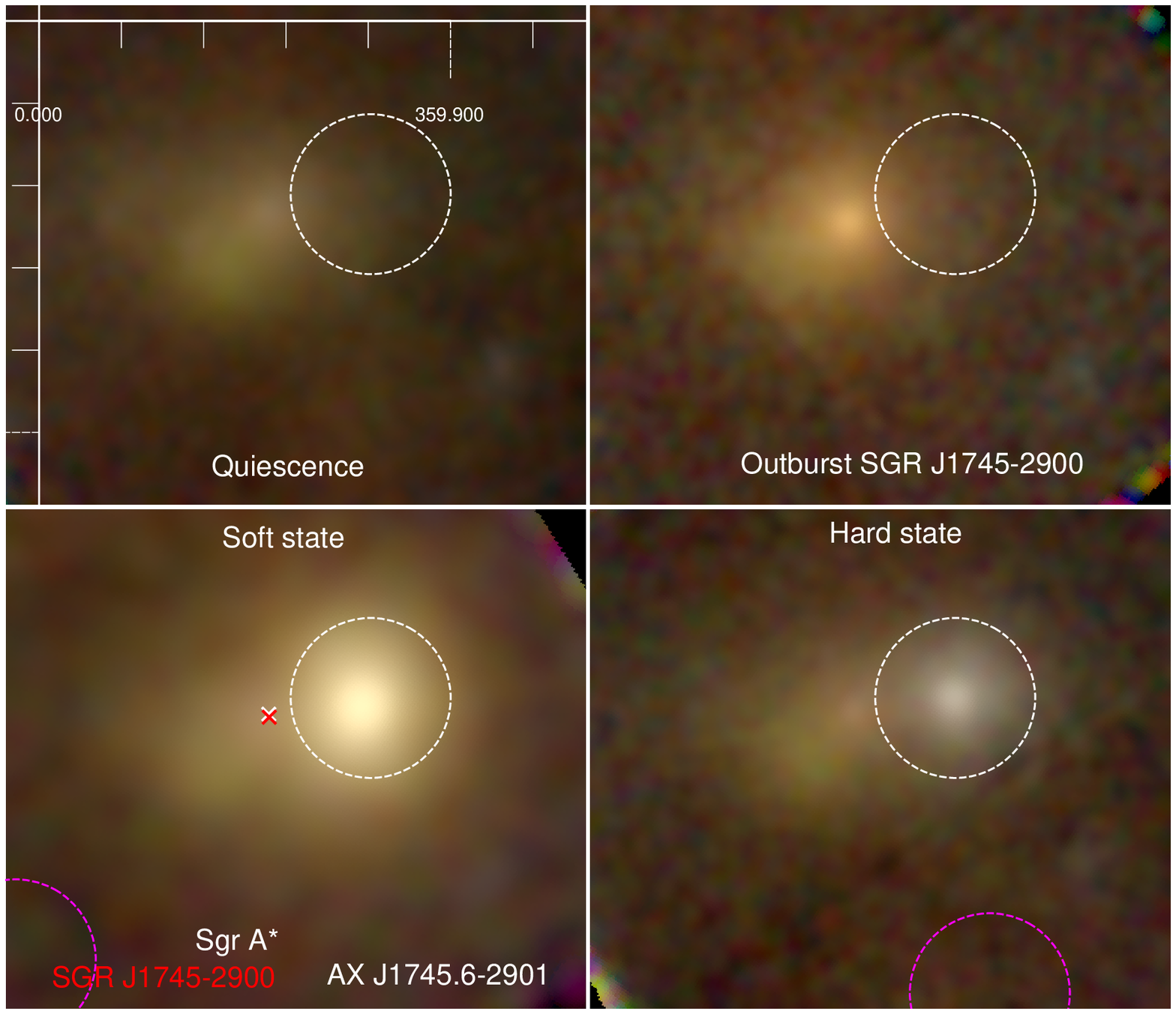}
\caption{\nustar\ exposure corrected RGB images of \axj\ 
(red, green and blue show the 3-6, 6-10 and 10-60 keV band, 
respectively). The white dashed circle shows the region used to 
extract the spectrum of \axj. The white and red crosses show the 
position of \sgras\ and \sgr, respectively. The top left panel shows the GC on 
2012-07-20, when no bright X-ray transient was observed and 
\axj\ was in quiescence. The scale is in Galactic coorodinates. 
The top right shows the GC X-ray emission 
on 2013-04-27, just before the outburst of \axj\ and around the 
peak of the outburst of \sgr. The bottom left and right panels show 
the X-ray emission on 2015-02-25 and 2014-09-27, while \axj\ 
was in the soft and hard state respectively. The magenta dashed 
circle shows the region used to extract the local background. 
The same logarithmic colour scale has been applied to all maps. }
\label{NuIma}
\end{figure}
The four panels in Fig. \ref{NuIma} show the \nustar\ RGB image 
of the regions around \axj. The top left panel shows the X-ray emission 
on 2012-07-20, when no bright X-ray transient was observed. 
Bright and diffuse X-ray emission is permeating the GC regions, 
producing a highly spatially variable background (see also Wang et 
al. 2002; Ponti et al. 2015; Mori et al. 2015; Fig. \ref{NuIma}). 
The right panel shows the X-ray emission on 2013-04-27, just 
after the start of the long outburst of \sgr, the magnetar that is 
located at $\sim1.45^{\prime}$ from \axj\ and only 
$\sim2.4^{\prime\prime}$ from \sgras\ (Rea et al. 2013; Mori et al. 
2013; Kaspi et al. 2014). 
We note that despite \sgr\ is lying outside of the extraction circle of \axj, 
a fraction of \sgr's photons enter into \axj's extraction region, 
polluting the spectra of \axj. \sgr's spectrum is very soft, it is best fit 
with $kT\sim0.7-0.9$~keV, an absorbed 1-10~keV flux 
of $F_{1-10}\sim(2-20)\times10^{-12}$~erg~cm$^{-2}$~s$^{-1}$, 
absorbed by a column density of neutral material of 
$N_H\sim1.5-2\times10^{23}$~cm$^{-2}$, exponentially decreasing 
over the period 2013-2014 (Mori et al. 2013; Rea et al. 2013; Kaspi 
et al. 2015; Coti-Zelati et al. 2015). 
The bottom panels of Fig. \ref{NuIma} show that the emission of 
\axj\ dominates over that of \sgr\ when it was in outburst (see 
the bottom left and right panels of Fig. \ref{NuIma}).

We note that during the observation performed on 2016 February 
$18^{th}$ (obsid: 90101022002), \swiftj, the accreting binary located 
at only $\sim1.5^{\prime}$ from \axj, was in outburst and brighter 
than \axj, therefore hampering a proper study of the spectrum 
of \axj\ (Degenaar et al. 2016; Ponti et al. 2016). 
For this reason, we excluded this observation from further analysis. 

\subsection{Background}
\label{Back}

To evaluate the importance on our results of the bright and highly spatially 
variable background emission, we extracted the background photons 
from several regions. 
\begin{figure}
\hspace{-0.6cm}
\includegraphics[height=0.509\textwidth,angle=-90]{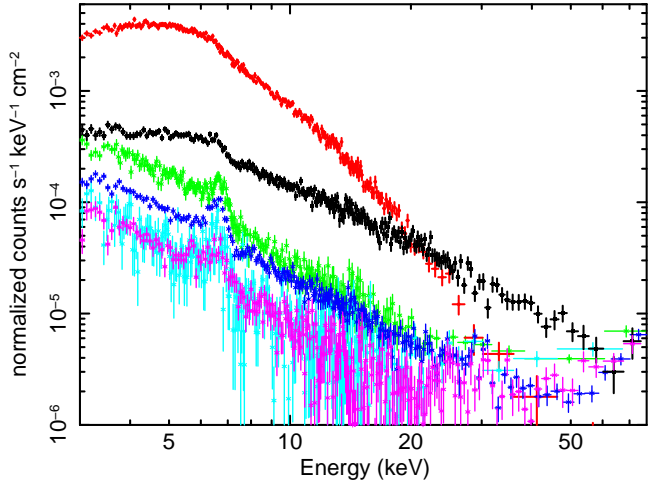}
\caption{\nustar\ spectra of \axj\ and of the background. 
The red dots and black squares show the spectra of \axj\ 
during the soft (2015-02-25) and hard state (2014-09-27), 
respectively. The blue asterisks show the background spectrum, 
from the same region used to extract the photons of \axj, 
when the source was quiescent and no X-ray transient was 
detected (2012-07-20). The green asterisks show the 
background from the same region, when \sgr\ was in outburst 
(2013-04-27). The cyan and magenta spectra show the local 
background spectra extracted from the magenta regions 
shown in Fig. \ref{NuIma}, during the soft and hard state, 
respectively. }
\label{NuBack}
\end{figure}

We initially selected the background during the longest \nustar\ 
exposure (accumulated in 2012-07-20; $\sim156.8$~ks), 
when \axj\ was in quiescence and no bright transient within a few 
arcmin from \sgras\ was active (see Fig. \ref{NuIma}). 
We extracted the background from the same region used for \axj's 
source emission (see \S \ref{NuDR}; a circular region with 
$70^{\prime\prime}$ radius centred on \axj). 
The blue asterisks in Fig. \ref{Back} show such background emission. 
We will refer to this background as {\it BackQuie}. 
We note that the {\it BackQuie} background dominates over the 
spectrum of \axj\ 
at energies above $\sim30-40$~keV during the soft (red) state and above 
$\sim60-70$ keV in the hard (black) state. Therefore, to avoid 
contamination from the background emission, we discarded the 
source photons above 40 and 70 keV during the soft and hard 
state, respectively. 

To estimate the maximum contamination of \sgr\ to the spectrum of 
\axj, we extracted a background spectrum from the same region used 
to extract \axj's photons, from an observation obtained on 2013-04-27, 
when \axj\ was quiescent and \sgr\ was at the peak of the X-ray 
outburst (Fig. \ref{NuIma}). We will refer to this background spectrum 
(green asterisks in Fig. \ref{Back}) as {\it BackSGR}. To extract a
quantitative measure of the emission induced by \sgr\ to the spectrum 
of \axj, we then fit the {\it BackSGR} spectrum, using {\it BackQuie} 
as background, with an absorbed black body model 
(with $N_H=1.6\times10^{23}$~cm$^{-2}$; Ponti et al. 2017). 
The best fit values are: $kT_{BB}=0.73$~keV and $r_{BB}=2.5$~km. 

We also extracted local background spectra from various regions 
around \axj. The cyan and magenta spectra in Fig. \ref{NuBack} show 
the local background from the observation accumulated on 2015-02-25 
and 2014-09-27, when \axj\ was in the soft and hard state, respectively 
(see the magenta regions shown in Fig. \ref{NuIma}). 
We chose these local background extraction regions by requiring them to be: 
i) located away from \axj\ and \sgras; ii) to be within the field of view; 
iii) to avoid X-ray transients as well as; iv) stray light and ghost rays. 

The comparison between the local and quiescent ({\it BackQuie}) 
backgrounds indicates that the latter are dominated by the diffuse 
GC emission, with only a small contribution from detector and 
particle background up to $\sim30-40$~keV (see Mori et al. 2015; 
Zhang et al. 2015 for more details). To correctly subtract the GC 
diffuse emission, we therefore chose to use {\it BackQuie} as 
background for the spectral analysis. 

\subsection{\nustar\ vs. \xmm\ low energy small mismatch}
\label{Xcal}

\subsubsection{Soft state}

\begin{figure}
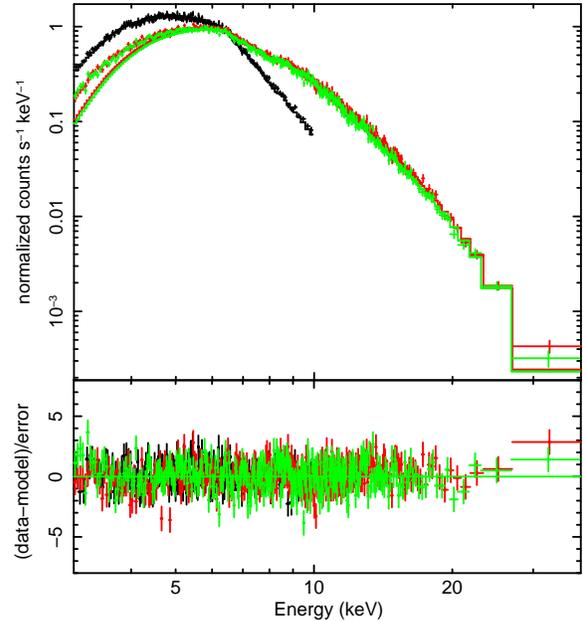

\includegraphics[height=0.45\textwidth,angle=-90]{Calib.ps}
\vspace{-0.1cm}

\includegraphics[height=0.45\textwidth,angle=-90]{crossdelS.ps}
\caption{{\it (Top panel:)} The black, red and green data shows the \xmm, \nustar\ 
FPMA and FPMB spectra obtained during the soft state simultaneous 
observation with the longest exposure (Tab. \ref{data} and \ref{TabNu}). 
The \xmm\ and \nustar\ spectra are fitted over the 3-10 keV and 5-40 
energy band. Clear deviations are observed at low energy when such 
model (best fitting the \xmm\ data) is extrapolated in the 3-5 keV range. 
{\it (Bottom panel:)} Residuals after fitting the spectra with model {\sc dbb-bb-rr}, 
allowing the column density of neutral material derived from 
the \nustar\ spectra to be different from the \xmm\ one 
(modified absorption). }
\label{crosscal}
\end{figure}
The black, red and green data in Fig. \ref{crosscal} show 
the \xmm, \nustar\ FPMA and FPMB spectra, respectively, obtained 
during the soft state observation with the longest simultaneous 
exposure (Tab. \ref{data} and \ref{TabNu}). 
We fit the \xmm\ and \nustar\ data with model {\sc dbb-bb-dl} (see \S 5.1.1).
The \xmm\ spectrum is corrected for the effects of dust scattering 
(see Jin et al. 2017). We fit the \xmm\ data within the 3-10~keV 
band and the \nustar\ ones in the 5-40 keV band. 
This model provides a reasonable fit of the data over the 
considered energy ranges ($\chi^2=1454.9$ for 1386 dof). 

Figure \ref{crosscal} shows that, the \nustar\ data over the 
$3-5$~keV band sit systematically above the extrapolation of 
the previously obtained best fit model. 
We correct \xmm\ for dust scattering effects using {\sc softabs}, 
but it is beyond the scope of the paper to do the same 
for \nustar, therefore we ignore NuSTAR below 5.5~keV, 
where dust scattering effects are prominent. 

\subsubsection{Hard state}

\begin{figure}
\includegraphics[height=0.45\textwidth,angle=-90]{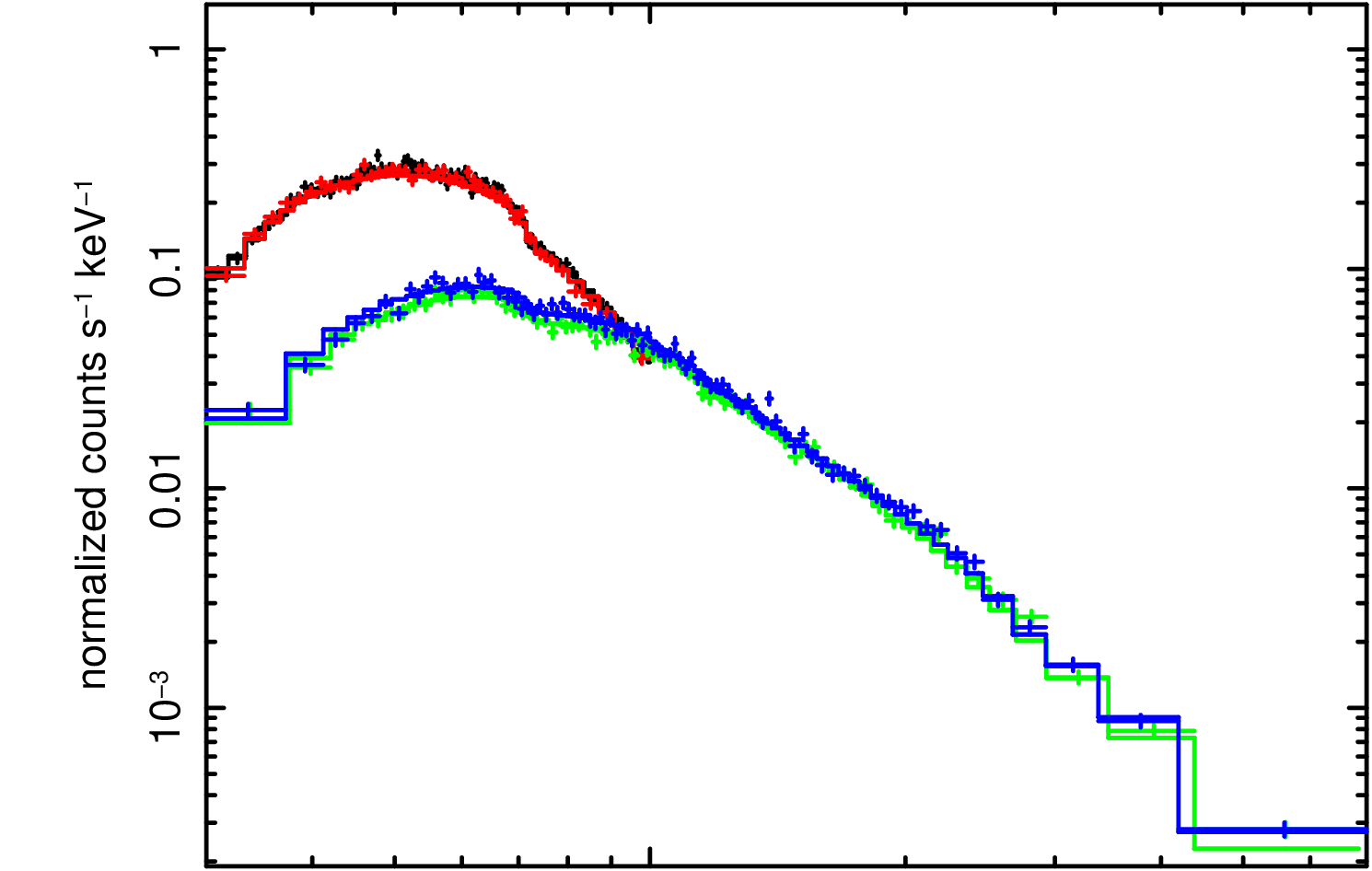}
\vspace{-0.1cm}

\includegraphics[height=0.45\textwidth,angle=-90]{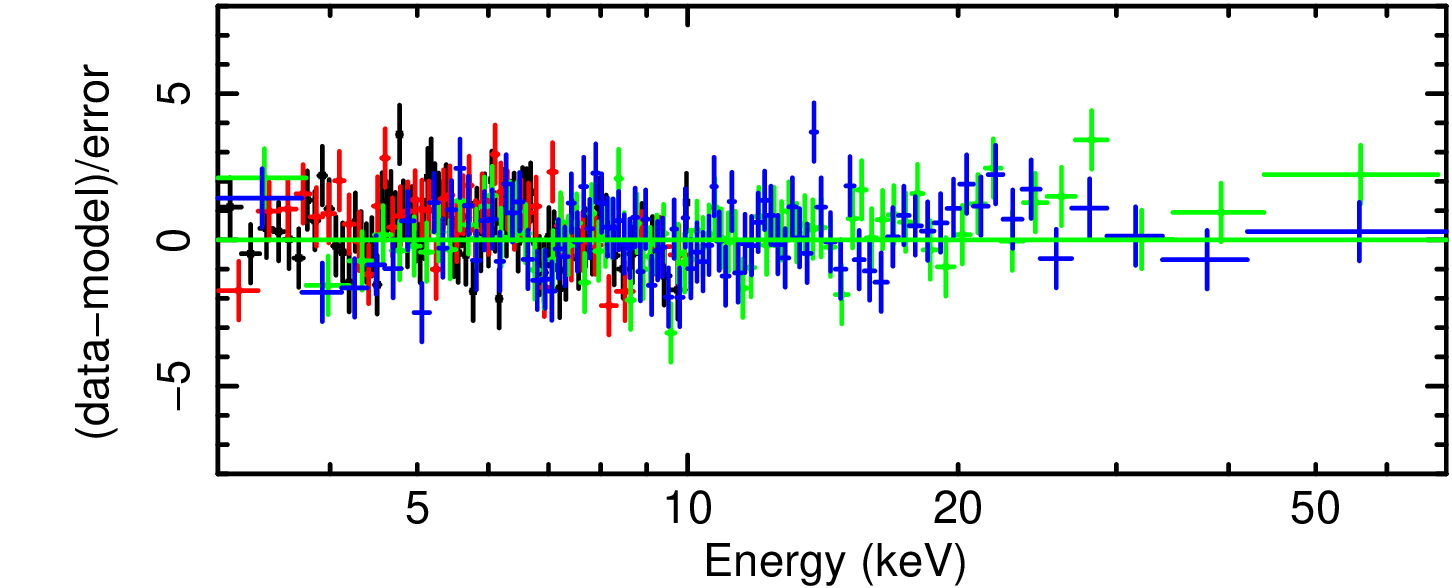}
\caption{{\it (Top panel:)} The black, red, green and blue data shows 
the two \xmm, \nustar\ FPMA and FPMB spectra obtained during 
the hard state simultaneous observation with the longest exposure 
(Tab. \ref{data} and \ref{TabNu}). 
The \xmm\ and \nustar\ spectra are fitted over the 3-10 keV and 5-40 
energy band. No residuals are observed at low energy when such 
model (best fitting the \xmm\ data) is extrapolated in the 3-5 keV range. 
{\it (Bottom panel:)} Residuals after fitting the spectra with model 
{\sc dbb-nth-rr}. }
\label{crosscalH}
\end{figure}
The black and red spectra in Fig. \ref{crosscalH} show the hard 
state \xmm\ spectra obtained on 2014-09-28 and -29, respectively, 
during the longest simultaneous \nustar\ exposure (Tab. \ref{data} 
and \ref{TabNu}). The green and blue spectra show the 
simultaneous FPMA and FPMB spectra. Because during the hard 
state there is no evidence for ionised absorption or a reflection 
component, we fit these hard state spectra with a simple power law 
component absorbed by neutral material.
The photon index observed by \nustar\ and 
within the two \xmm\ spectra are all consistent with each other 
(although the one measured by \nustar\ is slightly flatter), therefore 
we required the photon index to be the same for all spectra. 
We obtained a reasonable fit of the data ($\chi^2=2255.1$ for 
2118 dof). Fig. \ref{crosscalH} shows that this model  
reproduces not only the \xmm, but also the \nustar\ data 
even in the soft X-ray band, leaving no visible residuals. 

The small low energy mismatch detectable during the high flux 
soft state observations is not visible during the hard state observations, 
indicating that both \xmm\ and \nustar\ are very well cross-calibrated. 

\begin{table*}
\begin{center}
\small
\begin{tabular}{ c r c r r c r r c c c c c c c }
\hline
\hline
\multicolumn{2}{c}{\textbf{\emph{XMM-Newton}} } \\
 {\sc OBSID}         &    & Rev   &         START (UTC) & EXP  &CL EXP &STATE& F$_{3-6}$& F$_{6-10}$& F$_{8-10}$&Threshold\\
                             &    &          &                                & (ks)   & (ks)      &            &\multicolumn{3}{c}{($10^{-12}$~erg~cm$^{-2}$~s$^{-1}$)} & (ct~s$^{-1}$)   \\
\hline
{\it  0790180401} &    & 2970  & 2016-02-26 16:20:13  & 37.0 & 28.4 & S & 66.8 & 71.5   & 27.1 & 11/1.5/1.2/4.0/3.5/18       \\
{\bf 0743630901} & A & 2804  & 2015-04-02 08:59:19  & 36.6 &  4.5 &  S &101.9& 136.6 & 53.3 & 13/1.5/1.4/7.0/5.0/1.5      \\
{\bf 0743630801} & B & 2804  & 2015-04-01 08:34:19  & 27.0 & 20.1 & S & 96.3 & 121.7 & 46.4 & 13/1.5/1.3/6.2/4.9/1.1      \\
{\bf 0743630601} & C & 2787  & 2015-02-26 06:18:16  & 33.4 & 21.3 & S & 88.0 &101.7  & 37.7 & 12/1.5/1.4/5.9/4.4/5.0      \\
{\bf 0743630501} & D & 2711  & 2014-09-28 21:01:46  & 40.8 & 32.2 & H &   7.1 & 10.6   & 4.7   & 4.0/0.3/1.7/0.8/0.5/1.3     \\
{\bf 0743630401} & E & 2711  & 2014-09-27 17:30:23  & 34.5 & 18.6 & H &   7.3 & 10.5   & 4.6   & 3.5/0.2/1.5/0.8/0.5/2        \\ 
{\bf 0743630301} & F & 2697  & 2014-08-31 20:23:30  & 28.0 & 20.9 & H &   6.7 & 10.0   & 4.5   & 3.2/0.3/1.4/0.7/0.5/1.0     \\
{\bf 0743630201} & G & 2697  & 2014-08-30 19:20:01  & 35.0 & 22.4 & H &   7.5 & 10.7   & 4.7   & 3.5/0.3/1.4/0.75/0.55/1.5 \\
      0723410501  &     & 2621  & 2014-04-02 03:00:50  & 62.9 & 39.1 & H &   5.0 &   8.2   & 3.7   & 3.4/0.25/1.4/0.6/0.45/1.5 \\
      0723410401  &     & 2610  & 2014-03-10 14:10:43  & 57.0 & 38.5 & H &   5.3 &   8.3   & 3.6   & 3.2/0.25/1.4/0.58/0.4/0.8 \\
      0723410301  &     & 2605  & 2014-02-28 17:41:27  & 55.0 & 36.0 & H &   8.9 &  13.9  & 6.2   & 3.5/0.3/1.4/0.85/0.6/2.0   \\
\hline
\end{tabular}
\caption{
A list of all the {\it new} \xmm\ observations considered in this work. 
The columns of the table report the \xmm\ {\sc OBSID}, the \xmm\ revolution, 
the observation start date and time, the observation duration and the {\sc EPIC-pn} 
exposure time after cleaning, the source state (H=hard state; S=soft state). 
The following columns give the $3-6$, $6-10$ and $8-10$~keV 
observed (absorbed) fluxes in units of $10^{-12}$~erg~cm$^{-2}$~s$^{-1}$. 
The last column shows, in order, the count-rate thresholds applied to select bursting, 
eclipsing and intense dipping periods, the hard and soft count rates and the 
threshold to select out intense particle activity periods. A more exhaustive description 
of the data reduction and cleaning is provided in Section 2.1. With bold characters 
are reported the \xmm\ observations simultaneous with the \nustar\ ones. Italic 
characters indicate \xmm\ observations corrupted by very bright X-ray transients. }
\label{data}
\end{center}
\end{table*} 
\begin{table*}
\begin{center}
\small
\begin{tabular}{ c r c r r c r r c c c c c c c }
\hline
\hline
\multicolumn{2}{c}{\textbf{\emph{NuSTAR}} } \\
 {\sc OBSID}          &        & START (UTC) & EXP  & CL EXP&STATE& F$_{3-6}$& F$_{6-10}$& F$_{8-10}$&Threshold\\
                              &        &                        & (ks)   &    (ks)   &            &\multicolumn{3}{c}{($10^{-12}$~erg~cm$^{-2}$~s$^{-1}$)}& (ct~s$^{-1}$)   \\
\hline
{\it  90101022002} &        & 2016-02-18 22:26:08 & 33.8 &        & S &          & & & 5.5 1.0 \\  
      90101012002  &        & 2015-08-11 22:51:08 & 45.1 & 40.4 & S & 84.5 & 89.6 & 30.8 & 6.5 1.1  \\ 	
{\bf 30002002012} & A    & 2015-04-02 08:21:07 & 13.2 & 12.1 & S & 90.4 &116.4& 45.3 & 6.5 1.3  \\ 
{\bf 30002002010} & B    & 2015-04-01 06:31:07 & 14.4 & 11.9 & S & 80.8 & 90.6 & 33.7 & 5.0 1.2  \\ 
      30002002008   &       & 2015-03-31 04:41:07 & 25.7 & 21.8 & S & 73.6 & 77.3 & 27.9 & 5.0 0.9 \\ 
{\bf 30002002006} & C    & 2015-02-25 23:41:07 & 29.6 & 26.2 & S & 72.8 & 76.4 & 28.1 & 5.0 0.9  \\ 
{\bf 30002002004} & DE & 2014-09-27 17:31:07 & 67.8 & 59.5 & H & 4.9  & 7.9   & 3.6 & 2.1 0.07  \\ 	
{\bf 30002002002} & FG & 2014-08-30 19:46:07 & 60.3 & 50.6 & H & 4.8  & 7.7   & 3.6 & 2.1 0.09  \\ 	
      30001002010  &       & 2014-07-04 10:36:07 & 61.5 & 54.7 & H & 8.5  & 12.4 & 5.5 & 2.1 0.14  \\ 
      30001002008  &       & 2014-06-18 02:21:07 & 33.2 & 29.4 & H & 8.9  & 13.6 & 6.1 & 2.1 0.16  \\ 
      80002013026  &       & 2013-08-23 15:41:07 & 32.0 & 26.6 & S & 70.3 & 76.0 & 29.2 & 6.5 0.7  \\
      80002013024  &       & 2013-08-13 00:06:07 & 11.7 &  9.6  & S & 65.2 & 69.7 & 27.8 & 4.5 0.7  \\ 
      80002013022  &       & 2013-08-09 09:01:07 & 11.2 & 10.3 & S & 59.1 & 61.6 & 22.8 & 4.4 0.8  \\ 
      80002013020  &       & 2013-08-08 15:01:07 & 12.1 & 11.1 & S & 52.6 & 52.7 & 20.0 & 4.0 0.6  \\ 
      80002013018  &       & 2013-07-31 01:56:07 & 22.6 & 20.3 & H & 14.8 & 20.9 &  9.3 & 2.5 0.25  \\ 
\hline
\end{tabular}
\caption{
A list of all \nustar\ observations considered in this work. 
The columns of the table report the {\sc OBSID}, the observation start date, 
the observation duration and cleaned exposure, the source state (H=hard state; 
S=soft state). The following columns give the $3-6$, $6-10$ and $8-10$~keV 
observed (absorbed) fluxes in units of $10^{-12}$~erg~cm$^{-2}$~s$^{-1}$. 
The last column shows the count-rate thresholds applied to remove bursts, 
eclipses and dips. With bold characters are reported the observations 
simultaneous with the \xmm\ ones. Italic characters indicate observations 
corrupted by very bright X-ray transients. }
\label{TabNu}
\end{center}
\end{table*}

\end{document}